\documentclass[aps,prl,twocolumn,nopacs,superscriptaddress,amsmath,amssymb,groupedaddress]{revtex4}  
\usepackage{graphicx}  
\usepackage{dcolumn}   
\usepackage{bm}        
\usepackage{amssymb}
\usepackage{amsthm}
\usepackage{times}
\usepackage{bbold}
\usepackage{enumitem}
\usepackage{ulem}
\usepackage{color}
\makeatletter
\@ifundefined{textcolor}{}
{%
 \definecolor{BLACK}{gray}{0}
  \definecolor{GRAY}{gray}{0.6}
 \definecolor{WHITE}{gray}{1}
 \definecolor{RED}{rgb}{1,0,0}
 \definecolor{GREEN}{rgb}{0,1,0}
 \definecolor{BLUE}{rgb}{0,0,1}
 \definecolor{CYAN}{cmyk}{1,0,0,0}
 \definecolor{MAGENTA}{cmyk}{0,1,0,0}
 \definecolor{YELLOW}{cmyk}{0,0,1,0}
 \definecolor{ORANGE}{rgb}{0.9,0.5,0}
}

\newcommand{\jose}[1]{{\color{blue} #1}}

\newcommand{\rrangle}{\rangle\!\rangle} 
\newcommand{\llangle}{\langle\!\langle} 
\makeatother

\begin{document}
\let\vaccent=\v{
\global\long\def\gv#1{\ensuremath{\mbox{\boldmath\ensuremath{#1}}}}
\global\long\def\uv#1{\ensuremath{\mathbf{\hat{#1}}}}
\global\long\def\abs#1{\left| #1 \right|}
\global\long\def\avg#1{\left< #1 \right>}
\let\underdot=\d{
\global\long\def\dd#1#2{\frac{d^{2}#1}{d#2^{2}}}
\global\long\def\pd#1#2{\frac{\partial#1}{\partial#2}}
\global\long\def\pdd#1#2{\frac{\partial^{2}#1}{\partial#2^{2}}}
\global\long\def\pdc#1#2#3{\left( \frac{\partial#1}{\partial#2}\right)_{#3}}
\global\long\def\op#1{\hat{\mathrm{#1}}}
\global\long\def\ket#1{| #1 \rangle}
\global\long\def\bra#1{\langle #1 |}
\global\long\def\braket#1#2{\left< #1 \vphantom{#2}\right| \left. #2 \vphantom{#1}\right>}
\global\long\def\matrixel#1#2#3{\left< #1 \vphantom{#2#3}\right| #2 \left| #3 \vphantom{#1#2}\right>}
\global\long\def\av#1{\left\langle #1 \right\rangle }
 \global\long\def\com#1#2{\left[#1,#2\right]}
\global\long\def\acom#1#2{\left\{  #1,#2\right\}  }
\global\long\def\grad#1{\gv{\nabla} #1}
\let\divsymb=\div 
\global\long\def\div#1{\gv{\nabla} \cdot#1}
\global\long\def\curl#1{\gv{\nabla} \times#1}
\let\baraccent=\={

\title{Autonomous quantum error correction and quantum computation}


\author{Jos\'{e} Lebreuilly}
\affiliation{Department of Physics and Yale Quantum Institute,Yale University, New Haven, Connecticut 06520, USA}
\author{Kyungjoo Noh}
\thanks{This work was done before KN joined AWS Center for Quantum Computing.}
\affiliation{AWS Center for Quantum Computing, Pasadena, California 91125, USA}
\author{Chiao-Hsuan Wang}
\affiliation{Pritzker School of Molecular Engineering, University of Chicago, Chicago, Illinois 60637, USA}
\author{S. M. Girvin}
\affiliation{Department of Physics and Yale Quantum Institute,Yale University, New Haven, Connecticut 06520, USA}
\author{Liang Jiang}
\affiliation{Pritzker School of Molecular Engineering, University of Chicago, Chicago, Illinois 60637, USA}


\begin{abstract}
In this work, we present a general theoretical framework for the study of autonomously corrected quantum devices. First, we identify a necessary and sufficient revised version of the Knill-Laflamme conditions for the existence of an engineered Lindbladian providing protection against at most $c$ consecutive errors of natural dissipation, giving rise to an effective logical decoherence rate suppressed to order $c$. Moreover, we demonstrate that such engineered dissipation can be combined with generalized realizations of error-transparent Hamiltonians (ETH)  \cite{ETH_Vy,ETH,ETH_Rosenblum,PND_H,ETH_implem} in order to perform a quantum computation in the logical space while maintaining the same degree of suppression of decoherence. Finally, we introduce a formalism predicting with precision the emergent dynamics in the logical code space resulting from the interplay of natural, engineered dissipations sources and the generalized ETH.
\end{abstract}

\maketitle

\paragraph*{Introduction.}

Quantum error correction (QEC) plays a central role in the development of scalable quantum computers. While recent developments demonstrated the possibility to perform complex quantum information processing tasks with high fidelity in state-of-the-art architectures with 10-100 qubits \cite{Google_supr}, reliable large-scale quantum information processing still requires quantum error correction with efficient encoding and fault-tolerant implementation to suppress various practical imperfections \cite{nielsen_chuang, Campbell17}. The standard procedure for QEC relies on classical adaptive control, which measures the error syndrome and performs feedback with a classical controller. Recent experimental demonstration of QEC of bosonic encoding used classical adaptive control to reach the break-even point by suppressing the natural errors due to excitation loss \cite{Ofek16}. However, the performance of QEC with classical controllers is often limited by the readout errors \cite{Ofek16, HuL19}, decoherence from syndrome measurement \cite{Fluhmann19, Rosenblum18}, extra time delay and heating associated with classical feedback loop \cite{Reilly19, Campagne20}, as well as significant physical and computational resource overhead from the routine error recovery procedures  \cite{Steane03, Terhal15}. 

Alternatively, we may implement QEC \textit{without} classical adaptive control -- an approach called autonomous quantum error correction (AutoQEC) which has attracted increasing attention recently. Instead of measuring the error syndrome with a classical device, AutoQEC coherently processes the error syndrome and extract the entropy via engineered dissipation -- avoiding the measurement imperfection and overhead associated with classical feedback loop 
\footnote{If the syndrome extraction and engineered dissipation evolves continuous in time, AutoQEC becomes continuous time QEC (CTQEC) \cite{Paz98, Oreshkov_CTQEC, Hsu_CTQEC,Kyungjoo_AQEC}}.
AutoQEC has recently achieved efficient suppression of both cavity dephasing errors \cite{Lescanne20, Grimm20} and single-photon loss \cite{AQEC_gertler} using the bosonic cat encoding, which paves the way towards hardware-efficient QEC \cite{Mirrahimi_cat, Kapit_small_qubit, repetition_code_noise_bias, Puri20}. 
Theoretical investigations have shown that AutoQEC can improve the decoherence rate to $O(\kappa/R)$ at the logical level, where $\kappa$ is the natural error rate and $R\gg1$ is the ratio between the (good) engineered dissipation and (bad) natural error rates \cite{Paz98, Oreshkov_CTQEC, Hsu_CTQEC,Kyungjoo_AQEC}. 
In practice, however, it is still very challenging to achieve a very large R, because the strength of the engineered dissipation is limited by the driving power, various nonlinear processes, and finite energy gap \cite{Touzard18, Kyungjoo_AQEC,ZhangY19,Lescanne20}. 
Hence, it is desirable to use higher-order AutoQEC protocols, which can further improve logical decoherence rate to $O(\kappa/R^c)$ with $c>1$. In addition, from a scalability perspective (where small autonomously protected qubits would be used as basic elements of a larger architecture such as a topological code), it is also crucial to understand how logical gates can be applied while error correcting, and to assess the quality of the resulting logical qubits by quantifying systematically the post-AutoQEC residual logical error model.

In this work we perform a study of autonomously error corrected quantum devices and derive a fundamental result covering altogether AutoQEC  and autonomous error-corrected  approaches to quantum computation. Our formalism extends the Knill-Laflamme conditions to a continuous-time framework where natural and engineered dissipations are constantly acting on the system, and generalizes previous results \cite{Kyungjoo_AQEC} to quantum codes of arbitrary order $c\geq 1$ and arbitrary sets of jump error operators $F_{k}$. We prove that the Knill-Laflamme conditions are both necessary and sufficient conditions for AutoQEC. Under  those same conditions, we  show that autonomous error-corrected quantum computations with the same protection of order $c$ can be performed by combining the engineered dissipation with generalized versions of the standard error-transparent Hamiltonian \cite{ETH_Vy,ETH,ETH_Rosenblum,PND_H,ETH_implem}. In addition, we present a formalism allowing to assess with precision the post-AutoQEC dynamics in the restricted logical code space resulting from the interplay of natural and engineered dissipations, and the generalized ETH.

\begin{figure}
	\centering
	\includegraphics[width=1\columnwidth,clip]{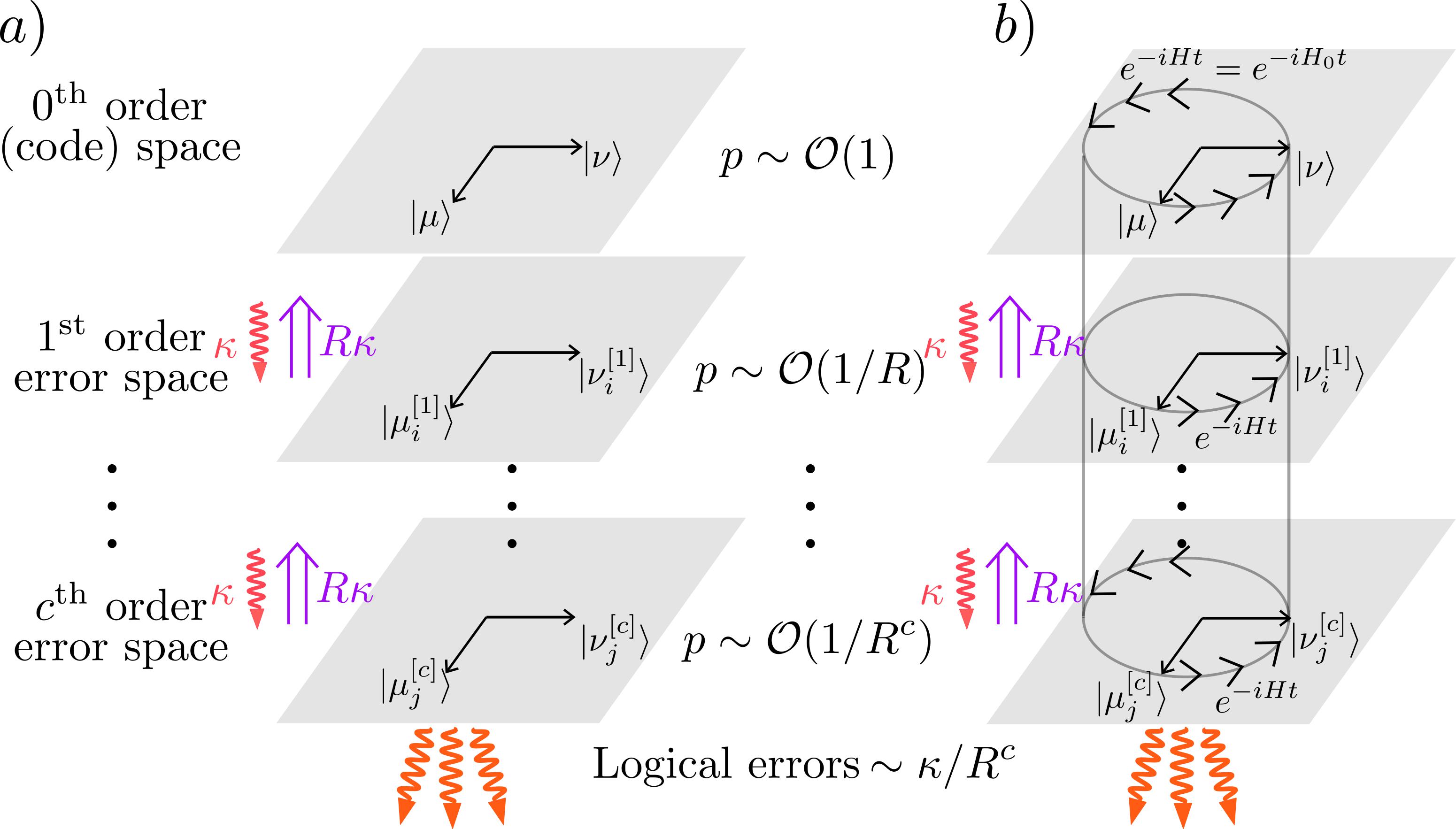}
	\caption{Main concepts underlying autonomous quantum error correction [Panel a)] and autonomous error-corrected quantum computation [Panel b)] at order $c$.\label{fig:sketch}}
\end{figure}
\paragraph*{Problem setup.}  
Let us have a Hilbert space $\mathcal{H}$ with some code subspace $\mathcal{C}\subset\mathcal{H}$, and some natural dissipation with jump operators $F_{k}$ acting on $\mathcal{H}$ with  the characteristic rate $\kappa$. One assumes that $d_{\mathcal{C}}=\textrm{dim}(\mathcal{C})$ and $d_{\mathcal{H}}=\textrm{dim}(\mathcal{H})$ are both finite and defines the code space projector $P_{\mathcal{C}}$ as well as its complementary projector $Q_{\mathcal{C}}=\mathbb{1}-P_{\mathcal{C}}$.  

With appropriate choice of logical subspace, we expect that the jump error operators $F_k$ cannot directly destroy the encoded quantum information. Instead, repeated action of the $F_k$ should generates a hierarchy of submanifolds corresponding to errors of incrementing weight $q$, as illustrated in Fig.~\ref{fig:sketch} a). For a quantum code of order $c\geq 1$, i.e., a code robust against at most $c$ errors, when the weight $q$ overcomes $c$ a logical error can however occur. Formally this structure should translate into Knill-Laflame conditions on an a certain error set, which is unknown at this stage and will be identified in a subsequent section. To obtain an effective  suppression of  decoherence of order $c$, a suitable set of additional jump operators $F_{\textrm{E},j}$ needs to be engineered, such that correctable error states are brought back to the code space without their underlying logical content being affected. If engineered dissipation has a characteristic rate $R\kappa$  (with $R$ dimensionless), in the strong recovery regime ($R\gg1$) the error submanifolds are expected to be strongly depleted, leading after kinetic equilibration  between  engineered and natural jumps to an occupancy of $q-$weighted error states scaling as $p_q\propto \mathcal{O}(1/R^q)$ for $q\leq c$ (see Fig.~\ref{fig:sketch} a)). Eventually logical errors occur when $c+1$ or more natural  errors are left unrecovered. As a consequence one expects the scaling
\begin{equation}
\kappa_{\textrm{eff}}\equiv \textrm{logical error rate} \underset{R\to+\infty}{=} \mathcal{O}\Big{(} \frac{\kappa}{R^{c}} \Big{)}, \label{def:scaling}
\end{equation}
for the resulting logical  decoherence  rate  in the code space.

Preferably, this suppression of decoherence should hold not only for an idle quantum state, i.e., for the purpose of performing AutoQEC, but also while performing some computation in the logical basis. In addition to the engineered dissipation, the latter requirement will involve the introduction of an error-transparent Hamiltonian \cite{ETH_Vy,ETH,ETH_Rosenblum,PND_H,ETH_implem}, or a generalization of an ETH. As sketched in Fig.~\ref{fig:sketch} b), in order to preserve  the scaling Eq.~(\ref{def:scaling}), such Hamiltonian should not only operate in the error-free  code space, but also couple  correctable error states  in a way which  does not  compromise the hierarchy of error submanifolds preexisting to the computation (e.g. by increasing the error weight),  and such that the correct logical operation has still been performed upon recovery by the engineered jumps $F_{\textrm{E},j}$. We formalize these concepts below.

\paragraph*{Definitions.} Let us consider the following Lindbladian time-continuous  evolution within the Hilbert space $\mathcal{H}$: 
\begin{equation}
\frac{d\rho}{dt} = \mathcal{L}\rho=\kappa\left(-i  g\left[ H,\rho\right]+R\mathcal{L}_{\rm{E}}(\rho)+\mathcal{L}_{\rm{n}}(\rho)\right)\label{eq:model}
\end{equation}
where $\mathcal{L}_{\textrm{n}}(\rho)=\sum_{k=1}^{N}D[F_{k}](\rho)$ and $\mathcal{L}_{\textrm{E}}(\rho)=\sum_{j}D[F_{\textrm{E},j}](\rho)$ are the Liouvillians associated with natural and  engineered dissipation,
and $D[A](\rho) \equiv A\rho A^{\dagger}-\frac{1}{2}\lbrace A^{\dagger}A,\rho \rbrace$. $H$ is the control Hamiltonian for logical operations. Apart from $\kappa$ with a dimension of a frequency, all introduced quantities are dimensionless. In particular the parameters $R$ and $g$ control respectively the relative speeds of engineered  dissipation and Hamiltonian dynamics compared to natural dissipation.  We introduce the useful notation $\mathcal{A}\otimes_d\mathcal{B}\equiv\textrm{span}\{\ket{\mu}\bra{\nu},\ket{\mu}\in \mathcal{A},\ket{\nu}\in \mathcal{B}\}$ and  the superoperator projector $\mathcal{P}_{\mathcal{C}}$ defined as $\mathcal{P}_{\mathcal{C}}\rho=P_{\mathcal{C}}\rho P_{\mathcal{C}}$ for all $\rho\in\mathcal{H}\otimes_d\mathcal{H}$. 
Since $\mathcal{L}_{\rm{E}}$ induces purely Lindbladian dynamics,  the limit $\mathcal{P}_{E}\equiv \textrm{lim}_{u\to\infty}\left[e^{\mathcal{L}_{\rm{E}}u}\right]$ is well-defined \cite{Lindbladian_dynamics,Lindbladian_dynamics_2} and is a projector ($\mathcal{P}_{E}^2=\mathcal{P}_{E}$).

First, let us consider the case $H=0$ of a protected quantum memory with no logical operation \footnote{Experimentally having $H=0$ might be achievable in a rotating frame where the qubit bare frequency has  been substracted.}. Assuming an initial condition $\rho(0)$ in the code space, due to natural dissipation the population leaks from the code space and occupies various error spaces (see Fig.~\ref{fig:sketch}a)) with a probability of order $\mathcal{O}(1/R)$. In order to assess the amount of logical information   in  $\rho(t)$ which has not been irreversibly erased by natural dissipation, it is therefore necessary to perform an information-preserving quantum recovery onto $\rho(t)$ back into the code space before comparing it to the initial state $\rho(0)$. One stresses that such recovery does not have to be explicitly implemented, but is only a mathematical tool to assess deviations at a finite time $t$. We now point out that if the engineered dissipation $\mathcal{L}_{\textrm{E}}$ is indeed able to protect the code space and perform autonomous error correction up to order $c$, then the CPTP projector $\mathcal{P}_{E}=\textrm{lim}_{u\to\infty}\left[e^{\mathcal{L}_{\rm{E}}u}\right]$ should be a fine hypothetical final recovery to assess such deviation. This leads us to the following definition: the engineered dissipation $\mathcal{L}_{\textrm{E}}$ performs \emph{ autonomous quantum error correction up  to order $c$} with respect to to the code space $\mathcal{C}$  and the natural dissipation $\mathcal{L}_{\textrm{n}}$  iff there exists $M>0$ such that  for all $ \rho(0)\in\mathcal{C}\otimes_d\mathcal{C}$, one has
\begin{equation}
\left\|\mathcal{P}_E\rho(t)-\rho(0)\right\|\leq M\|\rho(0)\|\frac{\kappa t}{R^c}\label{eq:definition_AutoQEC}
\end{equation}
for all times $t\geq 0$ and engineered dissipation strength $R\geq 0$, where $\rho(t)=e^{\mathcal{L}t}\rho(0)$. We  point out that since we work in a finite-dimensional space the choice of the norm $\|...\|$ for matrices in $\mathcal{H}\otimes_d\mathcal{H}$ only modifies the prefactor, which is not the focus of this investigation. 

We now move to the generic case $H\neq0$ where we want to perform a gate while preserving the same degree of accuracy. Consider some target dimensionless Hamiltonian $H_{0}=P_{\mathcal{C}}H_{0}P_{\mathcal{C}}$ acting only on  the logical space, we denote  $U_0(s)=\textrm{exp}[-iH_0 s]$ the unitary parametrized by the dimensionless time $s$. Accordingly, we say that the engineered dissipation $\mathcal{L}_{\rm{E}}$  and the Hamiltonian $H$ perform an \emph{autonomous error-corrected quantum computation up to order $c$} with respect to the code space $\mathcal{C}$, the target logical Hamiltonian $H_0$ and natural dissipation $\mathcal{L}_{\rm{n}}$, iff there exists two positive constants $M,g_0>0$ such that for any initial condition $\rho(0)\in\mathcal{C}\otimes_{d}\mathcal{C}$ one has
\begin{equation}
\left\|\mathcal{P}_{\rm{E}}\rho(t)-U_0(g  \kappa t)\rho(0)U_0(g\kappa t)^\dagger\right\| \leq M\|\rho(0)\|\frac{\kappa t}{R^c}\label{eq:definition_ETH_comp}
\end{equation}
for all $t,R\geq 0$, and all Hamiltonian coupling strengths $g$ satisfying $|g|\leq g_0 R$. We call any pair of $\mathcal{L}_{\rm{E}}$ and $H$ satisfying the latter condition respectively an \emph{error-correcting Liouvillian} and a \emph{generalized error-transparent Hamiltonian} of order $c$.

\paragraph{Error sets.} In connection to  the usual QEC formalism \cite{nielsen_chuang}  for discrete-time error channels with a Kraus representation, we expect an extension of Knill-Laflamme conditions to be satisfied on a certain error set (still to be identified) in order to  be able to perform AutoQEC in the continuous-time case. To facilitate the discussion, we define the zeroth- and first-order error sets as:
$\mathcal{E}^{[0]} \equiv \lbrace \mathbb{1} \rbrace$, $\mathcal{E}^{[1]} \equiv \lbrace F_{1},\cdots,F_{N} \rbrace$. At higher order we need to include errors from the no-jump evolution associated to the natural dissipation backaction Hamiltonian $H_{\rm{n,BA}}=\sum_{k=1}^{N}F_k^\dagger F_k$: for the second-order  error set, this  yields: $\mathcal{E}^{[2]}= \lbrace F_{k}F_l | k,l\in \lbrace 1,\cdots,N\rbrace \rbrace\cup\lbrace H_{\rm{n,BA}}\rbrace$. More generally, we construct higher-order error sets recursively as follows:
\begin{align}
	\mathcal{E}^{[n]} &\equiv \lbrace F_{k}E^{[n-1]}_{l} |E^{[n-1]}_{l}\in \mathcal{E}^{[n-1]}\textrm{ and } k\in \lbrace 1,\cdots,N\rbrace \rbrace
	\nonumber\\
	&\quad \cup  \lbrace H_{\rm{n,BA}}E^{[n-2]}_{l}|E^{[n-2]}_{l}\in \mathcal{E}^{[n-2]} \rbrace.
\end{align}

We employ some arbitrary labeling notation $\mathcal{E}^{[n]}=\{E_{j}^{[n]},1\leq  j \leq |\mathcal{E}^{[n]} | \}$ for the errors $E_{j}^{[n]}$ in $\mathcal{E}^{[n]}$.  Finally,  we define the error set up to order $n$: $\mathcal{E}^{[\sim n]}=\cup_{k=0}^{n} \mathcal{E}^{[k]}$. Below we formulate our main theorem on AutoQEC and autonomous error-corrected quantum computations, with its proof presented in \cite{SM}.

\paragraph{Theorem: existence of engineered dissipation and generalized error-transparent Hamiltonian.} 
Let us have some integer $c\geq 0$. The following statements are equivalent:
\begin{enumerate}[label=(\arabic*)]
	\item \emph{Knill-Laflamme:} The Knill-Laflamme condition is satisfied for the error set $\mathcal{E}^{[\sim c]}$.
	\item \emph{AutoQEC:} There exists a Liouvillian  $\mathcal{L}_{\textrm{E}}$ performing autonomous quantum error correction  up to order $c$ with respect to the code space $\mathcal{C}$ and the natural dissipation $\mathcal{L}_{\textrm{n}}$.
	\item \emph{Autonomous error-corrected quantum computations:} There exists an engineered dissipation  $\mathcal{L}_{\textrm{E}}$ satisfying the following property: for all logical Hamiltonians $H_0=P_{\mathcal{C}}H_{0}P_{\mathcal{C}}$, there exists a Hamiltonian $H$ such that   $\mathcal{L}_{\textrm{E}}$  and $H$ perform an autonomous error-corrected quantum computation up to order $c$  with respect to the code space $\mathcal{C}$, the target logical Hamiltonian $H_0$, and natural dissipation $\mathcal{L}_{\rm{n}}$.
\end{enumerate}
Despite the resulting equivalence between all these logical assertions, hypothesis (3) appears significantly stronger than hypothesis (2) as it ensures the possibility of performing any logical operation while error correcting. The equivalence between hypotheses (1) and (2) on the other hand is a direct extension of known results of QEC \cite{nielsen_chuang} to the context of continuous feedback and dissipation. Yet, there are some peculiarites emerging in this specific context: first, the ability to correct against  a  set of  quantum jump operations  $\{F_k\}$ does not guarantee to be able to correct against errors associated with superposition of error jump operators  $\alpha F_k+\beta F_j$. For example the modification of the backaction Hamiltonian $H_{\rm{n,BA}}$ generates new terms of  the  type $F_k^\dagger F_j$ which might not be included in the original error set, nor satisfy  the Knill-Laflamme conditions. Second,  although the backaction Hamiltonian $H_{\rm{n,BA}}$ acts as a  first-order term in the time evolution  of the density matrix $\rho(t)$,  it only counts as a  second-order error to  be corrected  ($H_{\rm{n,BA}}\in\mathcal{E}^{[2]}\backslash\mathcal{E}^{[1]}$).  The proof of the theorem, heavily relies on the following lemma.
\paragraph{Lemma: Properties of generalized error-transparent Hamiltonians and engineered dissipation.} Let us have some target Hamiltonian $H_{0}=P_{\mathcal{C}}H_{0}P_{\mathcal{C}}$, some engineered dissipation $\mathcal{L}_{\rm{E}}$ and generic Hamiltonian $H$.
We denote $\mathcal{H}$ (resp. $\mathcal{H}_0$ ) as the superoperator associated with the Hamiltonian evolution  $\mathcal{H}(\rho)=[H,\rho]$ (resp. $\mathcal{H}_0(\rho)=[H_0,\rho]$). The following statements are equivalent:
\begin{enumerate}[label=(\arabic*)]
	\item  $\mathcal{L}_{\textrm{E}}$  and $H$ perform an autonomous error-corrected quantum computation up to order $c$  with respect to the code space $\mathcal{C}$, the target logical Hamiltonian $H_0$ and natural dissipation $\mathcal{L}_{\rm{n}}$.
	\item for all sets of integers $k_1,k_2,k_3\in\mathbb{N}\times\mathbb{N}\times\{0,...,c\}$ one has 
	\begin{equation}
		\mathcal{P}_{\rm{E}}\mathcal{S}[\{\mathcal{H},k_1\},\{\mathcal{L}_{\rm{E}},k_2\},\{\mathcal{L}_{\rm{n}},k_3\}]\mathcal{P}_{\mathcal{C}}=
	\delta_{k_2,0}\delta_{k_3,0}\mathcal{H}_0^{k_1}\mathcal{P}_{\mathcal{C}}\label{eq:lemma}
\end{equation}
\end{enumerate}
The notation $\mathcal{S}\left(\{\mathcal{L}_{1},k_1\},\{\mathcal{L}_{2},k_2\},....,\{\mathcal{L}_{q},k_q\}\right)$ stands for the symmetrized product with unit prefactors of all possible orderings of $k_1$ powers of the $\mathcal{L}_{1}$, $k_2$ powers of the $\mathcal{L}_{2}$ up to $k_q$ powers of the $\mathcal{L}_{q}$. For example, $\mathcal{S}\left(\{\mathcal{L}_{1},2\},\{\mathcal{L}_{2},1\}\right)=\mathcal{L}_{1}^2\mathcal{L}_{2}+\mathcal{L}_{1}\mathcal{L}_{2}\mathcal{L}_{1}+\mathcal{L}_{2}\mathcal{L}_{1}^2$. As an alternative definition, symmetrized products $\mathcal{S}\left(\{\mathcal{L}_{1},k_1\},\{\mathcal{L}_{2},k_2\},....,\{\mathcal{L}_{q},k_q\}\right)$ can be defined as a generalization of the binomial expansion $(\sum_{i=1}^{q}\mathcal{L}_{i})^{K}=\sum_{k_1,...,k_q|\sum_{i}k_i =K}\mathcal{S}\left(\{\mathcal{L}_{1},k_1\},\{\mathcal{L}_{2},k_2\},....,\{\mathcal{L}_{q},k_q\}\right)$ in the non-commutative  case.

 The lemma tells us that the  final  recovery $\mathcal{P}_{\rm{E}}$ generated by engineered dissipation should not only be able to protect the code space against only natural dissipation events (of weight  at most $c$), but also against a larger class of processes intertwining natural dissipation and an  arbitrarily large number of engineered dissipation and Hamiltonian events. Interestingly, the final recovery is not required to recover arbitrary orderings between  these processes but only symmetrized terms $\mathcal{S}[\{\mathcal{H},k_1\},\{\mathcal{L}_{\rm{E}},k_2\},\{\mathcal{L}_{\rm{n}},k_3\}]$. This is because the evolution $\textrm{exp}\left[\kappa\left(-i  g\mathcal{H}+R\mathcal{L}_{\rm{E}}+\mathcal{L}_{\rm{n}}\right)t\right]$  resulting from time-independent couplings $g,R,\kappa$ only generates  symmetrized contributions when expanded in powers of $\mathcal{H}$, $\mathcal{L}_{\rm{E}}$, $\mathcal{L}_{\rm{n}}$ and $t$, and thus any asymmetric contributions  (which would be generated for time-dependent dynamics) are absent here and do not need to be corrected. 
While the complete proofs of the theorem and the  Lemma can be found in \cite{SM}, we detail below an explicit construction of engineered dissipation and Hamiltonian satisfying the Lemma conditions and thus performing AutoQEC and error-corrected autonomus computations.

\paragraph{Explicit construction of an engineered dissipation and Hamiltonian for error-corrected computations.}
We assume that the Knill-Laflamme condition is satisfied for the  error set $\mathcal{E}^{[\sim c]}$. Let us have $\ket{\mu}\in\mathcal{C}$. We construct an orthogonal basis of correctable error states as follows:  
$\mu^{\mathcal{E}}\equiv( |\mu\rangle =|\mu_{1 = p_{0}}^{[0]}\rangle , |\mu_{1}^{[1]}\rangle, ...,|\mu_{p_{1}}^{[1]}\rangle,..., |\mu_{1}^{[c]}\rangle,...,|\mu_{p_{c}}^{[c]}\rangle )=\textrm{G.S.}(|\mu\rangle, E_{1}^{[1]}|\mu_{1}\rangle,... E_{|\mathcal{E}^{[1]}|}^{[1]}|\mu_{1}\rangle,...,E_{1}^{[c]}|\mu_{1}\rangle, ..., E_{|\mathcal{E}^{[c]}|}^{[c]}|\mu_{1}\rangle)$. The notation $\textrm{G.S.}(v_i)_{i\in I}$ stands for the uniquely defined orthonormal vector family generated by the Gram-Schmidt algorithm using as input the vector family $(v_i)_{i\in I}$ defined over an ordered ensemble $I$. For all $k\in[0,...,c]$, the vectors $|\mu_{j}^{[k]}\rangle$ represent on orthogonal basis of the $k$-th order error states generated from  the logical $\ket{\mu}\in\mathcal{C}$which are orthogonal to all lower-order error states. The Knill-Laflamme conditions being satisfied, we can show for all $k\in\{0,..,c\}$ that the number  $p_{k}$ of generated states $\{|\mu_{i}^{[k]}\rangle\}_{1\leq  i\leq p_k}$ is independent of $\ket{\mu}\in\mathcal{C}$, and that for two choices  $\ket{\mu}\perp \ket{\nu}\in\mathcal{C}$, the vectors of $\mu^{\mathcal{E}}$ are orthogonal to the vectors of $\nu^{\mathcal{E}}$. Finally, given a basis $\{\ket{\mu_i},1\leq i \leq d_{\mathcal{C}}\}$ of the code space, one defines the space $\mathcal{H}_{\rm{res}}$ of residual states as $\mathcal{H}_{\rm{res}}=\left(\textrm{span}\big{\lbrace} \cup_{i=1}^{d_{\mathcal{C}}} \mu_i^{\mathcal{E}} \big\rbrace\right)^{\perp} = \textrm{span}\lbrace |\phi_{1}\rangle,\cdots,|\phi_{q_{\max} }\rangle \rbrace$
where $q_{\max} = d_{\mathcal{H}} - d_{\mathcal{C}}\sum_{n=0}^{c}p_{n}$. An engineered dissipation $\mathcal{L}_{\rm{E}}$ and generalized error-transparent Hamiltonian $H$ satisfying the condition (2) of the Lemma and thus performing an autonomous error-corrected quantum computation up to order $c$ are then constructed as follows:
\begin{eqnarray}
\mathcal{L}_{\textrm{E}} &=& \sum_{n=1}^{c}\sum_{i_{n}=1}^{p_{n}} D[ F_{\textrm{E},i_{n}}^{[n]} ] + \sum_{q=1}^{q_{\max}} D[F_{\textrm{E},q}^{[\textrm{res}]}]\label{eq:explicit_L},\\
H&=&\sum_{j,k=1}^{d_{\mathcal{C}}}\sum_{n=0}^{c}\sum_{i_{n}=1}^{p_{n}} \langle\mu_j|H_0|\mu_k\rangle|\mu_{j,i_{n}}^{[n]}\rangle\langle\mu_{k,i_{n}}^{[n]}|.\label{eq:explicit_H}
\end{eqnarray}
The jump operators associated with engineered dissipation are given by
\begin{align}
F_{\textrm{E},i_{n}}^{[n]} &= \sum_{j=1}^{d_{\mathcal{C}}}|\mu_{j}\rangle\langle \mu_{j,i_{n}}^{[n]}| \quad \textrm{for}\,\, 1 \le n\le c \,\, \textrm{and} \,\, 1\le i_{n}\le p_{n} , 
\nonumber\\
F_{\textrm{E},q}^{[\textrm{res}]} &= |\Phi_{q}\rangle \langle \phi_{q}| \quad \textrm{for} \,\, 1\le q \le q_{\max}, 
\end{align}
and $|\Phi_{q}\rangle$ is any (normalized) state in $\mathcal{C}$.  
\paragraph{Generalized error-transparent Hamiltonians.} 
Interestingly, in order to perform correctly an error-corrected computation of order $c$, an Hamiltonian does not necessarily need to be error-transparent  in the standard sense, which has been originally formulated as a requirement of commutation with errors \cite{ETH_Vy,ETH}. While,  the specific construction in Eq.~(\ref{eq:explicit_H}) indeed accidentally satisfies both conditions $HEP_{\mathcal{C}}=EH_0P_{\mathcal{C}}$ and $[H,E]P_{\mathcal{C}}=0$ for all the error operators  $E\in\mathcal{E}^{[\sim c]}$  of our error set, from our Lemma, we see that a generalized ETH $H$ can be defined as any Hamiltonian satisfying the condition of Eq.~(\ref{eq:lemma}). Such condition admits more general solutions: in \cite{SM} we  outline in particular examples where a generalized ETH $H$ does not commute with errors, preserve the error syndrome, nor satisfy $HP_{\mathcal{C}}= H_0P_{\mathcal{C}}$ (in the latter case $H$ was found to not stabilize the code space). 

\paragraph{Logical space dynamics.}
For practical purposes it is often useful to acquire a more precise knowledge of the resulting decoherence dynamics of the corrected logical  qubit. Consider some engineered dissipation $\mathcal{L}_{\rm{E}}$ and Hamiltonian $H$ satisfying Eq.~(\ref{eq:definition_ETH_comp}). In particular, according to the Lemma, the recovery projector satisfies $\mathcal{P}_{\rm{E}}\mathcal{P}_{\mathcal{C}}=\mathcal{P}_{\mathcal{C}}$. For simplicity, we will also assume that the steady-states of $\mathcal{L}_{\rm{E}}$  are  all contained in $\mathcal{C}$, which translates as $\mathcal{P}_{\mathcal{C}}\mathcal{P}_{\rm{E}}=\mathcal{P}_{\rm{E}}$. Note that  our explicit construction for  $\mathcal{L}_{\rm{E}}$  in Eq.~(\ref{eq:explicit_L}) does satisfy this  condition. 

Under these conditions, in \cite{SM} we present the derivation of an effective master equation governing the dynamics of $\mathcal{P}_{\rm{E}}\rho(t)$.  Our formalism is based on Nakajima-Zwanzig projection operator techniques \cite{Breuer_projective}:  assuming an  initial condition $\rho(0)$ in the code space, $\mathcal{P}_{\rm{E}}\rho(t)$ follows the exact non-Markovian dynamics
$\partial_t\mathcal{P}_{\rm{E}}\rho=\mathcal{P}_{\rm{E}}\mathcal{L}\mathcal{P}_{\rm{E}}+\int_0^t  d\tau \Sigma(\tau)\mathcal{P}_{\rm{E}}\rho(t-\tau)$,
where $\Sigma(\tau)=\mathcal{P}_{\rm{E}}\mathcal{L}\mathcal{Q}_{\rm{E}}\textrm{exp}[\mathcal{Q}_{\rm{E}}\mathcal{L}\mathcal{Q}_{\rm{E}}\tau]\mathcal{Q}_{\rm{E}}\mathcal{L}\mathcal{P}_{\rm{E}}$ is the Nakajima-Zwanzig memory kernel, and $\mathcal{Q}_{\rm{E}}\equiv \mathcal{I}-\mathcal{P}_{\rm{E}}$ is the complementary projector to $\mathcal{P}_{\rm{E}}$.  After highlighting a strong time scale separation between the relaxation dynamics of the memory kernel $\Sigma(\tau)$ and of the projected density matrix $\mathcal{P}_{\rm{E}}\rho(t)$ occuring respectively over $t\sim 1/(R\kappa)$ and $t\sim  R^c/\kappa$ as soon as $R\gg1$  \cite{SM}, we derive the following time-local master equation $\partial_t\mathcal{P}_{\rm{E}}\rho(t)=[-i\kappa g\mathcal{H}_0+\mathcal{L}_{\rm{eff}}]\mathcal{P}_{\rm{E}}\rho(t)+[\partial_t\mathcal{P}_{\rm{E}}\rho(t)]_{\rm{corr}}$, where
\begin{equation}
\mathcal{L}_{\rm{eff}}=\int_{0}^{+\infty} d\tau\left[\Sigma(\tau)e^{i g\kappa\mathcal{H}_0\tau}\right]\mathcal{P}_{\mathcal{C}}.
\end{equation}
Meanwhile, the full density matrix satisfies $\rho(t)=\mathcal{T}\mathcal{P}_{\rm{E}}\rho(t)+\delta\rho(t)$, where
$ \mathcal{T}=\mathcal{I}+\int_0^{+\infty} d\tau e^{\mathcal{Q}_{\rm{E}}\mathcal{L}\mathcal{Q}_{\rm{E}}\tau}\mathcal{Q}_{\rm{E}}\mathcal{L}e^{i g\kappa\mathcal{H}_0\tau}\mathcal{P}_{\mathcal{\rm{C}}}$.
 As desired, the effective Liouvillian superoperator $\mathcal{L}_{\rm{eff}}$ only acts within the code space: $ \mathcal{L}_{\rm{eff}}=\mathcal{P}_{\mathcal{C}}\mathcal{L}_{\rm{eff}}=\mathcal{L}_{\rm{eff}}\mathcal{P}_{\mathcal{C}}$. The full density matrix $\rho(t)$ contains components in error spaces due to the application of  $\mathcal{T}$ on the projected matrix $\mathcal{P}_{\rm{E}}\rho(t)$. 
 The corrections to these effective dynamics are shown to satisfy the following upperbounds
 $\|[\partial_t\mathcal{P}_{\rm{E}}\rho(t)]_{\rm{corr}}\|\leq \kappa\left[K/R^{2c+1}+\tilde{K}e^{-\beta\kappa Rt}/R^c\right]$, $\|\delta\rho(t)\|\leq Ae^{-R\kappa B t}/R +C/R^{c+2}$ for some  $A,B,C,K>0$. These bounds  hold for all $\kappa,t,R\geq 0$, $|g|\leq \tilde{g}_0R$  for some constant $\tilde{g}_0>0$. We conclude that for large $R$ the derived estimate can become arbitrarily precise, and that the increase in precision occurs faster for error-correcting codes of larger order $c$. 
 
 
The  expressions of the integrals in $\mathcal{L}_{\rm{eff}}$ and $ \mathcal{T}$ are reported in the general case in the Supplemental Material \cite{SM}. We present here the corresponding  results in the case $H=H_0=0$ without computation in the logical basis: 
\begin{equation}
\mathcal{L}_{\rm{eff}}=\kappa\sum_{k=c+1}^{+\infty}\left(\frac{-1}{R}\right)^{k-1}\mathcal{P}_{\rm{E}}\left(\mathcal{L}_{\rm{n}}\mathcal{L}_{\rm{E}}^{\ast}\right)^{k-1}\mathcal{L}_{\rm{n}}\mathcal{P}_{\mathcal{C}}\label{eq:effective-liouvillian_auto_QEC}
\end{equation}
with correspondingly
$\mathcal{T}=\sum_{n=0}^{\infty}\frac{(-1)^k}{R^k}\left(\mathcal{L}_{\rm{E}}^{\ast}\mathcal{L}_{\rm{n}}\right)^{k}\mathcal{P}_{\mathcal{C}}$. The quantity $\mathcal{L}_{\rm{E}}^{\ast}=-\int_0^{+\infty}du [e^{\mathcal{L}_{\rm{E}}u}\mathcal{Q}_{\rm{E}}]$ is well-defined as $\mathcal{Q}_{\rm{E}}$ projects onto the relaxation modes of $\mathcal{L}_{\rm{E}}$, and is a pseudo inverse of $\mathcal{L}_{\rm{E}}$.  Along with along $\mathcal{P}_{\rm{E}}$, $\mathcal{L}_{\rm{E}}^{\ast}$ possesses a relatively simple analytical expression within our explicit construction of Eqs.~(\ref{eq:explicit_L}-\ref{eq:explicit_H})  (see \cite{SM}). As expected  intuitively from the AutoQEC of order $c$ hypothesis, the summation in Eq.~(\ref{eq:effective-liouvillian_auto_QEC}) initiates with terms involving at least $k=c+1$ powers of natural  dissipation, and  the effective Liouvillian $\mathcal{L}_{\rm{eff}}$ is indeed suppressed as $1/R^c$ for a large $R$. This scaling is demonstrated in  \cite{SM} to persist in the presence of non-vanishing Hamiltonians $H$ and $H_0$.

  \paragraph*{Concluding remarks and outlooks}
 We have developed an extensive mathematical framework for autonomously error-corrected quantum devices. 
 In particular, our study features a proof in the time-continuous context of the  equivalence between Knill-Laflamme conditions  and the possibility to perform both autonomous quantum error correction and error-corrected quantum computations to  an arbitrary order $c\geq1$. Our study encompasses all types of codes and  Markovian error models. An important generalization of these results  regards time-dependent and non-Markovian  dissipation sources.  Future  work will also investigate approximated quantum error correction and noise-biased logical qubits in connection to the relevant applications to bosonic quantum computation.
 
\begin{acknowledgments}\textit{Acknowledgements.}  

We acknowledge support from the ARO (W911NF-18-1-0020, W911NF-18-1-0212), ARO MURI (W911NF-16-1-0349), AFOSR MURI (FA9550-19-1-0399), DOE (DE-SC0019406), NSF (EFMA-1640959, OMA-1936118, EEC-1941583), NTT Research, and the Packard Foundation (2013-39273).

\end{acknowledgments}
\bibliographystyle{apsrev4-1}
\bibliography{Bibliography}

\begin{thebibliography}{34}%
\makeatletter
\providecommand \@ifxundefined [1]{%
 \@ifx{#1\undefined}
}%
\providecommand \@ifnum [1]{%
 \ifnum #1\expandafter \@firstoftwo
 \else \expandafter \@secondoftwo
 \fi
}%
\providecommand \@ifx [1]{%
 \ifx #1\expandafter \@firstoftwo
 \else \expandafter \@secondoftwo
 \fi
}%
\providecommand \natexlab [1]{#1}%
\providecommand \enquote  [1]{``#1''}%
\providecommand \bibnamefont  [1]{#1}%
\providecommand \bibfnamefont [1]{#1}%
\providecommand \citenamefont [1]{#1}%
\providecommand \href@noop [0]{\@secondoftwo}%
\providecommand \href [0]{\begingroup \@sanitize@url \@href}%
\providecommand \@href[1]{\@@startlink{#1}\@@href}%
\providecommand \@@href[1]{\endgroup#1\@@endlink}%
\providecommand \@sanitize@url [0]{\catcode `\\12\catcode `\$12\catcode
  `\&12\catcode `\#12\catcode `\^12\catcode `\_12\catcode `\%12\relax}%
\providecommand \@@startlink[1]{}%
\providecommand \@@endlink[0]{}%
\providecommand \url  [0]{\begingroup\@sanitize@url \@url }%
\providecommand \@url [1]{\endgroup\@href {#1}{\urlprefix }}%
\providecommand \urlprefix  [0]{URL }%
\providecommand \Eprint [0]{\href }%
\providecommand \doibase [0]{http://dx.doi.org/}%
\providecommand \selectlanguage [0]{\@gobble}%
\providecommand \bibinfo  [0]{\@secondoftwo}%
\providecommand \bibfield  [0]{\@secondoftwo}%
\providecommand \translation [1]{[#1]}%
\providecommand \BibitemOpen [0]{}%
\providecommand \bibitemStop [0]{}%
\providecommand \bibitemNoStop [0]{.\EOS\space}%
\providecommand \EOS [0]{\spacefactor3000\relax}%
\providecommand \BibitemShut  [1]{\csname bibitem#1\endcsname}%
\let\auto@bib@innerbib\@empty
\bibitem [{\citenamefont {Vy}\ \emph {et~al.}(2013)\citenamefont {Vy},
  \citenamefont {Xiaoting~Wang},\ and\ \citenamefont {Jacobs}}]{ETH_Vy}%
  \BibitemOpen
  \bibfield  {author} {\bibinfo {author} {\bibfnamefont {O.}~\bibnamefont
  {Vy}}, \bibinfo {author} {\bibfnamefont {X.}~\bibnamefont {Xiaoting~Wang}}, \
  and\ \bibinfo {author} {\bibfnamefont {K.}~\bibnamefont {Jacobs}},\
  }\href@noop {} {\bibfield  {journal} {\bibinfo  {journal} {N. J. Phys.}\
  }\textbf {\bibinfo {volume} {15}},\ \bibinfo {pages} {053002} (\bibinfo
  {year} {2013})}\BibitemShut {NoStop}%
\bibitem [{\citenamefont {Kapit}(2018)}]{ETH}%
  \BibitemOpen
  \bibfield  {author} {\bibinfo {author} {\bibfnamefont {E.}~\bibnamefont
  {Kapit}},\ }\href@noop {} {\bibfield  {journal} {\bibinfo  {journal} {Phys.
  Rev. Lett.}\ }\textbf {\bibinfo {volume} {120}},\ \bibinfo {pages} {050503}
  (\bibinfo {year} {2018})}\BibitemShut {NoStop}%
\bibitem [{\citenamefont {Rosenblum}\ \emph
  {et~al.}(2018{\natexlab{a}})\citenamefont {Rosenblum}, \citenamefont
  {Reinhold}, \citenamefont {Mirrahimi}, \citenamefont {Jiang}, \citenamefont
  {Frunzio},\ and\ \citenamefont {Schoelkopf}}]{ETH_Rosenblum}%
  \BibitemOpen
  \bibfield  {author} {\bibinfo {author} {\bibfnamefont {S.}~\bibnamefont
  {Rosenblum}}, \bibinfo {author} {\bibfnamefont {P.}~\bibnamefont {Reinhold}},
  \bibinfo {author} {\bibfnamefont {M.}~\bibnamefont {Mirrahimi}}, \bibinfo
  {author} {\bibfnamefont {L.}~\bibnamefont {Jiang}}, \bibinfo {author}
  {\bibfnamefont {L.}~\bibnamefont {Frunzio}}, \ and\ \bibinfo {author}
  {\bibfnamefont {R.~J.}\ \bibnamefont {Schoelkopf}},\ }\href@noop {}
  {\bibfield  {journal} {\bibinfo  {journal} {Science}\ }\textbf {\bibinfo
  {volume} {361}},\ \bibinfo {pages} {266} (\bibinfo {year}
  {2018}{\natexlab{a}})}\BibitemShut {NoStop}%
\bibitem [{\citenamefont {Wang}\ \emph {et~al.}(2020)\citenamefont {Wang},
  \citenamefont {Noh}, \citenamefont {Lebreuilly}, \citenamefont {Girvin},\
  and\ \citenamefont {Jiang}}]{PND_H}%
  \BibitemOpen
  \bibfield  {author} {\bibinfo {author} {\bibfnamefont {C.-H.}\ \bibnamefont
  {Wang}}, \bibinfo {author} {\bibfnamefont {K.}~\bibnamefont {Noh}}, \bibinfo
  {author} {\bibfnamefont {J.}~\bibnamefont {Lebreuilly}}, \bibinfo {author}
  {\bibfnamefont {S.~M.}\ \bibnamefont {Girvin}}, \ and\ \bibinfo {author}
  {\bibfnamefont {L.}~\bibnamefont {Jiang}},\ }\href@noop {} {\bibfield
  {journal} {\bibinfo  {journal} {arXiv:2009.07855}\ } (\bibinfo {year}
  {2020})}\BibitemShut {NoStop}%
\bibitem [{\citenamefont {Ma}\ \emph {et~al.}(2020)\citenamefont {Ma},
  \citenamefont {Xu}, \citenamefont {Mu}, \citenamefont {Cai}, \citenamefont
  {Hu}, \citenamefont {Wang}, \citenamefont {Pan}, \citenamefont {Wang},
  \citenamefont {Song}, \citenamefont {Zou},\ and\ \citenamefont
  {Sun}}]{ETH_implem}%
  \BibitemOpen
  \bibfield  {author} {\bibinfo {author} {\bibfnamefont {Y.}~\bibnamefont
  {Ma}}, \bibinfo {author} {\bibfnamefont {Y.}~\bibnamefont {Xu}}, \bibinfo
  {author} {\bibfnamefont {X.}~\bibnamefont {Mu}}, \bibinfo {author}
  {\bibfnamefont {W.}~\bibnamefont {Cai}}, \bibinfo {author} {\bibfnamefont
  {L.}~\bibnamefont {Hu}}, \bibinfo {author} {\bibfnamefont {W.}~\bibnamefont
  {Wang}}, \bibinfo {author} {\bibfnamefont {X.}~\bibnamefont {Pan}}, \bibinfo
  {author} {\bibfnamefont {H.}~\bibnamefont {Wang}}, \bibinfo {author}
  {\bibfnamefont {Y.~P.}\ \bibnamefont {Song}}, \bibinfo {author}
  {\bibfnamefont {C.~L.}\ \bibnamefont {Zou}}, \ and\ \bibinfo {author}
  {\bibfnamefont {L.}~\bibnamefont {Sun}},\ }\href@noop {} {\bibfield
  {journal} {\bibinfo  {journal} {Nat. Phys.}\ }\textbf {\bibinfo {volume}
  {16}},\ \bibinfo {pages} {827} (\bibinfo {year} {2020})}\BibitemShut
  {NoStop}%
\bibitem [{\citenamefont {Arute~et al.}(2019)}]{Google_supr}%
  \BibitemOpen
  \bibfield  {author} {\bibinfo {author} {\bibfnamefont {F.}~\bibnamefont
  {Arute~et al.}},\ }\href@noop {} {\bibfield  {journal} {\bibinfo  {journal}
  {Nature}\ }\textbf {\bibinfo {volume} {574}},\ \bibinfo {pages} {505}
  (\bibinfo {year} {2019})}\BibitemShut {NoStop}%
\bibitem [{\citenamefont {Nielsen}\ and\ \citenamefont
  {Chuang}(2000)}]{nielsen_chuang}%
  \BibitemOpen
  \bibfield  {author} {\bibinfo {author} {\bibfnamefont {M.~A.}\ \bibnamefont
  {Nielsen}}\ and\ \bibinfo {author} {\bibfnamefont {I.~L.}\ \bibnamefont
  {Chuang}},\ }\href@noop {} {\emph {\bibinfo {title} {Quantum Computation and
  Quantum Information}}}\ (\bibinfo  {publisher} {Cambridge University Press},\
  \bibinfo {year} {2000})\BibitemShut {NoStop}%
\bibitem [{\citenamefont {Campbell}\ \emph {et~al.}(2017)\citenamefont
  {Campbell}, \citenamefont {Terhal},\ and\ \citenamefont
  {Vuillot}}]{Campbell17}%
  \BibitemOpen
  \bibfield  {author} {\bibinfo {author} {\bibfnamefont {E.~T.}\ \bibnamefont
  {Campbell}}, \bibinfo {author} {\bibfnamefont {B.~M.}\ \bibnamefont
  {Terhal}}, \ and\ \bibinfo {author} {\bibfnamefont {C.}~\bibnamefont
  {Vuillot}},\ }\href@noop {} {\bibfield  {journal} {\bibinfo  {journal}
  {Nature}\ }\textbf {\bibinfo {volume} {549}},\ \bibinfo {pages} {172}
  (\bibinfo {year} {2017})}\BibitemShut {NoStop}%
\bibitem [{\citenamefont {Ofek}\ \emph {et~al.}(2016)\citenamefont {Ofek},
  \citenamefont {P.}, \citenamefont {Heeres}, \citenamefont {Reinhold},
  \citenamefont {Leghtas}, \citenamefont {Vlastakis}, \citenamefont {Liu},
  \citenamefont {Frunzio}, \citenamefont {Girvin}, \citenamefont {Jiang},
  \citenamefont {Mirrahimi}, \citenamefont {Devoret},\ and\ \citenamefont
  {Schoelkopf}}]{Ofek16}%
  \BibitemOpen
  \bibfield  {author} {\bibinfo {author} {\bibfnamefont {N.}~\bibnamefont
  {Ofek}}, \bibinfo {author} {\bibfnamefont {A.}~\bibnamefont {P.}}, \bibinfo
  {author} {\bibfnamefont {R.}~\bibnamefont {Heeres}}, \bibinfo {author}
  {\bibfnamefont {P.}~\bibnamefont {Reinhold}}, \bibinfo {author}
  {\bibfnamefont {Z.}~\bibnamefont {Leghtas}}, \bibinfo {author} {\bibfnamefont
  {B.}~\bibnamefont {Vlastakis}}, \bibinfo {author} {\bibfnamefont
  {Y.}~\bibnamefont {Liu}}, \bibinfo {author} {\bibfnamefont {L.}~\bibnamefont
  {Frunzio}}, \bibinfo {author} {\bibfnamefont {S.~M.}\ \bibnamefont {Girvin}},
  \bibinfo {author} {\bibfnamefont {L.}~\bibnamefont {Jiang}}, \bibinfo
  {author} {\bibfnamefont {M.}~\bibnamefont {Mirrahimi}}, \bibinfo {author}
  {\bibfnamefont {M.~H.}\ \bibnamefont {Devoret}}, \ and\ \bibinfo {author}
  {\bibfnamefont {R.~J.}\ \bibnamefont {Schoelkopf}},\ }\href@noop {}
  {\bibfield  {journal} {\bibinfo  {journal} {Nature}\ }\textbf {\bibinfo
  {volume} {536}},\ \bibinfo {pages} {441} (\bibinfo {year}
  {2016})}\BibitemShut {NoStop}%
\bibitem [{\citenamefont {Hu}\ \emph {et~al.}(2019)\citenamefont {Hu},
  \citenamefont {Ma}, \citenamefont {Cai}, \citenamefont {Mu}, \citenamefont
  {Xu}, \citenamefont {Wang}, \citenamefont {Wu}, \citenamefont {Wang},
  \citenamefont {Song}, \citenamefont {Zou}, \citenamefont {Girvin},
  \citenamefont {Duan},\ and\ \citenamefont {Sun}}]{HuL19}%
  \BibitemOpen
  \bibfield  {author} {\bibinfo {author} {\bibfnamefont {L.}~\bibnamefont
  {Hu}}, \bibinfo {author} {\bibfnamefont {Y.}~\bibnamefont {Ma}}, \bibinfo
  {author} {\bibfnamefont {W.}~\bibnamefont {Cai}}, \bibinfo {author}
  {\bibfnamefont {X.}~\bibnamefont {Mu}}, \bibinfo {author} {\bibfnamefont
  {Y.}~\bibnamefont {Xu}}, \bibinfo {author} {\bibfnamefont {W.}~\bibnamefont
  {Wang}}, \bibinfo {author} {\bibfnamefont {Y.}~\bibnamefont {Wu}}, \bibinfo
  {author} {\bibfnamefont {H.}~\bibnamefont {Wang}}, \bibinfo {author}
  {\bibfnamefont {Y.~P.}\ \bibnamefont {Song}}, \bibinfo {author}
  {\bibfnamefont {C.~L.}\ \bibnamefont {Zou}}, \bibinfo {author} {\bibfnamefont
  {S.~M.}\ \bibnamefont {Girvin}}, \bibinfo {author} {\bibfnamefont {L.~M.}\
  \bibnamefont {Duan}}, \ and\ \bibinfo {author} {\bibfnamefont
  {L.}~\bibnamefont {Sun}},\ }\href@noop {} {\bibfield  {journal} {\bibinfo
  {journal} {Nature Physics}\ }\textbf {\bibinfo {volume} {15}},\ \bibinfo
  {pages} {503} (\bibinfo {year} {2019})}\BibitemShut {NoStop}%
\bibitem [{\citenamefont {Flühmann}\ \emph {et~al.}(2019)\citenamefont
  {Flühmann}, \citenamefont {Nguyen}, \citenamefont {Marinelli}, \citenamefont
  {Negnevitsky}, \citenamefont {Mehta},\ and\ \citenamefont
  {Home}}]{Fluhmann19}%
  \BibitemOpen
  \bibfield  {author} {\bibinfo {author} {\bibfnamefont {C.}~\bibnamefont
  {Flühmann}}, \bibinfo {author} {\bibfnamefont {T.~L.}\ \bibnamefont
  {Nguyen}}, \bibinfo {author} {\bibfnamefont {M.}~\bibnamefont {Marinelli}},
  \bibinfo {author} {\bibfnamefont {V.}~\bibnamefont {Negnevitsky}}, \bibinfo
  {author} {\bibfnamefont {K.}~\bibnamefont {Mehta}}, \ and\ \bibinfo {author}
  {\bibfnamefont {J.~P.}\ \bibnamefont {Home}},\ }\href@noop {} {\bibfield
  {journal} {\bibinfo  {journal} {Nature}\ }\textbf {\bibinfo {volume} {566}},\
  \bibinfo {pages} {513} (\bibinfo {year} {2019})}\BibitemShut {NoStop}%
\bibitem [{\citenamefont {Rosenblum}\ \emph
  {et~al.}(2018{\natexlab{b}})\citenamefont {Rosenblum}, \citenamefont
  {Reinhold}, \citenamefont {Mirrahimi}, \citenamefont {Jiang}, \citenamefont
  {Frunzio},\ and\ \citenamefont {Schoelkopf}}]{Rosenblum18}%
  \BibitemOpen
  \bibfield  {author} {\bibinfo {author} {\bibfnamefont {S.}~\bibnamefont
  {Rosenblum}}, \bibinfo {author} {\bibfnamefont {P.}~\bibnamefont {Reinhold}},
  \bibinfo {author} {\bibfnamefont {M.}~\bibnamefont {Mirrahimi}}, \bibinfo
  {author} {\bibfnamefont {L.}~\bibnamefont {Jiang}}, \bibinfo {author}
  {\bibfnamefont {L.}~\bibnamefont {Frunzio}}, \ and\ \bibinfo {author}
  {\bibfnamefont {R.~J.}\ \bibnamefont {Schoelkopf}},\ }\href@noop {}
  {\bibfield  {journal} {\bibinfo  {journal} {Science}\ }\textbf {\bibinfo
  {volume} {361}},\ \bibinfo {pages} {266} (\bibinfo {year}
  {2018}{\natexlab{b}})}\BibitemShut {NoStop}%
\bibitem [{\citenamefont {Reilly}(2019)}]{Reilly19}%
  \BibitemOpen
  \bibfield  {author} {\bibinfo {author} {\bibfnamefont {D.~J.}\ \bibnamefont
  {Reilly}},\ }\href@noop {} {\bibfield  {journal} {\bibinfo  {journal} {2019
  IEEE International Electron Devices Meeting (IEDM)}\ ,\ \bibinfo {pages}
  {31.7}} (\bibinfo {year} {2019})}\BibitemShut {NoStop}%
\bibitem [{\citenamefont {Campagne-Ibarcq}\ \emph {et~al.}(2020)\citenamefont
  {Campagne-Ibarcq}, \citenamefont {Eickbusch}, \citenamefont {Touzard},
  \citenamefont {Zalys-Geller}, \citenamefont {Frattini}, \citenamefont
  {Sivak}, \citenamefont {Reinhold}, \citenamefont {Puri}, \citenamefont
  {Shankar}, \citenamefont {Schoelkopf}, \citenamefont {Frunzio}, \citenamefont
  {Mirrahimi},\ and\ \citenamefont {Devoret}}]{Campagne20}%
  \BibitemOpen
  \bibfield  {author} {\bibinfo {author} {\bibfnamefont {P.}~\bibnamefont
  {Campagne-Ibarcq}}, \bibinfo {author} {\bibfnamefont {A.}~\bibnamefont
  {Eickbusch}}, \bibinfo {author} {\bibfnamefont {S.}~\bibnamefont {Touzard}},
  \bibinfo {author} {\bibfnamefont {E.}~\bibnamefont {Zalys-Geller}}, \bibinfo
  {author} {\bibfnamefont {N.~E.}\ \bibnamefont {Frattini}}, \bibinfo {author}
  {\bibfnamefont {V.~V.}\ \bibnamefont {Sivak}}, \bibinfo {author}
  {\bibfnamefont {P.}~\bibnamefont {Reinhold}}, \bibinfo {author}
  {\bibfnamefont {S.}~\bibnamefont {Puri}}, \bibinfo {author} {\bibfnamefont
  {S.}~\bibnamefont {Shankar}}, \bibinfo {author} {\bibfnamefont {R.~J.}\
  \bibnamefont {Schoelkopf}}, \bibinfo {author} {\bibfnamefont
  {L.}~\bibnamefont {Frunzio}}, \bibinfo {author} {\bibfnamefont
  {M.}~\bibnamefont {Mirrahimi}}, \ and\ \bibinfo {author} {\bibfnamefont
  {M.~H.}\ \bibnamefont {Devoret}},\ }\href@noop {} {\bibfield  {journal}
  {\bibinfo  {journal} {Nature}\ }\textbf {\bibinfo {volume} {584}},\ \bibinfo
  {pages} {368} (\bibinfo {year} {2020})}\BibitemShut {NoStop}%
\bibitem [{\citenamefont {Steane}(2003)}]{Steane03}%
  \BibitemOpen
  \bibfield  {author} {\bibinfo {author} {\bibfnamefont {A.~M.}\ \bibnamefont
  {Steane}},\ }\href@noop {} {\bibfield  {journal} {\bibinfo  {journal}
  {Physical Review A}\ }\textbf {\bibinfo {volume} {68}},\ \bibinfo {pages}
  {042322} (\bibinfo {year} {2003})}\BibitemShut {NoStop}%
\bibitem [{\citenamefont {Terhal}(2015)}]{Terhal15}%
  \BibitemOpen
  \bibfield  {author} {\bibinfo {author} {\bibfnamefont {B.~M.}\ \bibnamefont
  {Terhal}},\ }\href@noop {} {\bibfield  {journal} {\bibinfo  {journal}
  {Reviews of Modern Physics}\ }\textbf {\bibinfo {volume} {87}},\ \bibinfo
  {pages} {307} (\bibinfo {year} {2015})}\BibitemShut {NoStop}%
\bibitem [{\citenamefont {Lescanne}\ \emph {et~al.}(2020)\citenamefont
  {Lescanne}, \citenamefont {Villiers}, \citenamefont {Peronnin}, \citenamefont
  {Sarlette}, \citenamefont {Delbecq}, \citenamefont {Huard}, \citenamefont
  {Kontos}, \citenamefont {Mirrahimi},\ and\ \citenamefont
  {Leghtas}}]{Lescanne20}%
  \BibitemOpen
  \bibfield  {author} {\bibinfo {author} {\bibfnamefont {R.}~\bibnamefont
  {Lescanne}}, \bibinfo {author} {\bibfnamefont {M.}~\bibnamefont {Villiers}},
  \bibinfo {author} {\bibfnamefont {T.}~\bibnamefont {Peronnin}}, \bibinfo
  {author} {\bibfnamefont {A.}~\bibnamefont {Sarlette}}, \bibinfo {author}
  {\bibfnamefont {M.}~\bibnamefont {Delbecq}}, \bibinfo {author} {\bibfnamefont
  {B.}~\bibnamefont {Huard}}, \bibinfo {author} {\bibfnamefont
  {T.}~\bibnamefont {Kontos}}, \bibinfo {author} {\bibfnamefont
  {M.}~\bibnamefont {Mirrahimi}}, \ and\ \bibinfo {author} {\bibfnamefont
  {Z.}~\bibnamefont {Leghtas}},\ }\href@noop {} {\bibfield  {journal} {\bibinfo
   {journal} {Nature Physics}\ }\textbf {\bibinfo {volume} {16}},\ \bibinfo
  {pages} {509} (\bibinfo {year} {2020})}\BibitemShut {NoStop}%
\bibitem [{\citenamefont {Grimm}\ \emph {et~al.}(2020)\citenamefont {Grimm},
  \citenamefont {Frattini}, \citenamefont {Puri}, \citenamefont {Mundhada},
  \citenamefont {Touzard}, \citenamefont {Mirrahimi}, \citenamefont {Girvin},
  \citenamefont {Shankar},\ and\ \citenamefont {Devoret}}]{Grimm20}%
  \BibitemOpen
  \bibfield  {author} {\bibinfo {author} {\bibfnamefont {A.}~\bibnamefont
  {Grimm}}, \bibinfo {author} {\bibfnamefont {N.~E.}\ \bibnamefont {Frattini}},
  \bibinfo {author} {\bibfnamefont {S.}~\bibnamefont {Puri}}, \bibinfo {author}
  {\bibfnamefont {S.~O.}\ \bibnamefont {Mundhada}}, \bibinfo {author}
  {\bibfnamefont {S.}~\bibnamefont {Touzard}}, \bibinfo {author} {\bibfnamefont
  {M.}~\bibnamefont {Mirrahimi}}, \bibinfo {author} {\bibfnamefont {S.~M.}\
  \bibnamefont {Girvin}}, \bibinfo {author} {\bibfnamefont {S.}~\bibnamefont
  {Shankar}}, \ and\ \bibinfo {author} {\bibfnamefont {M.~H.}\ \bibnamefont
  {Devoret}},\ }\href@noop {} {\bibfield  {journal} {\bibinfo  {journal}
  {Nature}\ }\textbf {\bibinfo {volume} {584}},\ \bibinfo {pages} {205}
  (\bibinfo {year} {2020})}\BibitemShut {NoStop}%
\bibitem [{\citenamefont {Gertler}\ \emph {et~al.}(2020)\citenamefont
  {Gertler}, \citenamefont {Baker}, \citenamefont {Li}, \citenamefont {Shirol},
  \citenamefont {Koch},\ and\ \citenamefont {Wang}}]{AQEC_gertler}%
  \BibitemOpen
  \bibfield  {author} {\bibinfo {author} {\bibfnamefont {J.~M.}\ \bibnamefont
  {Gertler}}, \bibinfo {author} {\bibfnamefont {B.}~\bibnamefont {Baker}},
  \bibinfo {author} {\bibfnamefont {J.}~\bibnamefont {Li}}, \bibinfo {author}
  {\bibfnamefont {S.}~\bibnamefont {Shirol}}, \bibinfo {author} {\bibfnamefont
  {J.}~\bibnamefont {Koch}}, \ and\ \bibinfo {author} {\bibfnamefont
  {C.}~\bibnamefont {Wang}},\ }\href@noop {} {\bibfield  {journal} {\bibinfo
  {journal} {arXiv:2004.09322}\ } (\bibinfo {year} {2020})}\BibitemShut
  {NoStop}%
\bibitem [{\citenamefont {Mirrahimi}\ \emph {et~al.}(2014)\citenamefont
  {Mirrahimi}, \citenamefont {Leghtas}, \citenamefont {Albert}, \citenamefont
  {Touzard}, \citenamefont {Schoelkopf}, \citenamefont {Jiang},\ and\
  \citenamefont {Devoret}}]{Mirrahimi_cat}%
  \BibitemOpen
  \bibfield  {author} {\bibinfo {author} {\bibfnamefont {M.}~\bibnamefont
  {Mirrahimi}}, \bibinfo {author} {\bibfnamefont {Z.}~\bibnamefont {Leghtas}},
  \bibinfo {author} {\bibfnamefont {V.~V.}\ \bibnamefont {Albert}}, \bibinfo
  {author} {\bibfnamefont {S.}~\bibnamefont {Touzard}}, \bibinfo {author}
  {\bibfnamefont {R.~J.}\ \bibnamefont {Schoelkopf}}, \bibinfo {author}
  {\bibfnamefont {L.}~\bibnamefont {Jiang}}, \ and\ \bibinfo {author}
  {\bibfnamefont {M.~H.}\ \bibnamefont {Devoret}},\ }\href@noop {} {\bibfield
  {journal} {\bibinfo  {journal} {New Journal of Physics}\ }\textbf {\bibinfo
  {volume} {16}},\ \bibinfo {pages} {045014} (\bibinfo {year}
  {2014})}\BibitemShut {NoStop}%
\bibitem [{\citenamefont {Kapit}(2016)}]{Kapit_small_qubit}%
  \BibitemOpen
  \bibfield  {author} {\bibinfo {author} {\bibfnamefont {E.}~\bibnamefont
  {Kapit}},\ }\href@noop {} {\bibfield  {journal} {\bibinfo  {journal} {Phys.
  Rev. Lett.}\ }\textbf {\bibinfo {volume} {116}},\ \bibinfo {pages} {150501}
  (\bibinfo {year} {2016})}\BibitemShut {NoStop}%
\bibitem [{\citenamefont {Guillaud}\ and\ \citenamefont
  {Mirrahimi}(2019)}]{repetition_code_noise_bias}%
  \BibitemOpen
  \bibfield  {author} {\bibinfo {author} {\bibfnamefont {J.}~\bibnamefont
  {Guillaud}}\ and\ \bibinfo {author} {\bibfnamefont {M.}~\bibnamefont
  {Mirrahimi}},\ }\href@noop {} {\bibfield  {journal} {\bibinfo  {journal}
  {Phys. Rev. X}\ }\textbf {\bibinfo {volume} {9}},\ \bibinfo {pages} {041053}
  (\bibinfo {year} {2019})}\BibitemShut {NoStop}%
\bibitem [{\citenamefont {Puri}\ \emph {et~al.}(2020)\citenamefont {Puri},
  \citenamefont {St-Jean}, \citenamefont {Gross}, \citenamefont {Grimm},
  \citenamefont {Frattini}, \citenamefont {Iyer}, \citenamefont {Krishna},
  \citenamefont {Touzard}, \citenamefont {Jiang}, \citenamefont {Blais},
  \citenamefont {Flammia},\ and\ \citenamefont {Girvin}}]{Puri20}%
  \BibitemOpen
  \bibfield  {author} {\bibinfo {author} {\bibfnamefont {S.}~\bibnamefont
  {Puri}}, \bibinfo {author} {\bibfnamefont {L.}~\bibnamefont {St-Jean}},
  \bibinfo {author} {\bibfnamefont {J.~A.}\ \bibnamefont {Gross}}, \bibinfo
  {author} {\bibfnamefont {A.}~\bibnamefont {Grimm}}, \bibinfo {author}
  {\bibfnamefont {N.~E.}\ \bibnamefont {Frattini}}, \bibinfo {author}
  {\bibfnamefont {P.~S.}\ \bibnamefont {Iyer}}, \bibinfo {author}
  {\bibfnamefont {A.}~\bibnamefont {Krishna}}, \bibinfo {author} {\bibfnamefont
  {S.}~\bibnamefont {Touzard}}, \bibinfo {author} {\bibfnamefont
  {L.}~\bibnamefont {Jiang}}, \bibinfo {author} {\bibfnamefont
  {A.}~\bibnamefont {Blais}}, \bibinfo {author} {\bibfnamefont {S.~T.}\
  \bibnamefont {Flammia}}, \ and\ \bibinfo {author} {\bibfnamefont {S.~M.}\
  \bibnamefont {Girvin}},\ }\href@noop {} {\bibfield  {journal} {\bibinfo
  {journal} {Science Advances}\ }\textbf {\bibinfo {volume} {6}},\ \bibinfo
  {pages} {eaay5901} (\bibinfo {year} {2020})}\BibitemShut {NoStop}%
\bibitem [{\citenamefont {Paz}\ and\ \citenamefont {Zurek}(1998)}]{Paz98}%
  \BibitemOpen
  \bibfield  {author} {\bibinfo {author} {\bibfnamefont {J.~P.}\ \bibnamefont
  {Paz}}\ and\ \bibinfo {author} {\bibfnamefont {W.~H.}\ \bibnamefont
  {Zurek}},\ }\href@noop {} {\bibfield  {journal} {\bibinfo  {journal}
  {Proceedings of the Royal Society of London. Series A: Mathematical, Physical
  and Engineering Sciences}\ }\textbf {\bibinfo {volume} {454}},\ \bibinfo
  {pages} {355} (\bibinfo {year} {1998})}\BibitemShut {NoStop}%
\bibitem [{\citenamefont {Oreshkov}(2013)}]{Oreshkov_CTQEC}%
  \BibitemOpen
  \bibfield  {author} {\bibinfo {author} {\bibfnamefont {O.}~\bibnamefont
  {Oreshkov}},\ }\href@noop {} {\emph {\bibinfo {title} {Quantum Error
  Correction}}}\ (\bibinfo  {publisher} {Cambridge University Press},\ \bibinfo
  {year} {2013})\ Chap.~\bibinfo {chapter} {8}, p.\ \bibinfo {pages}
  {201}\BibitemShut {NoStop}%
\bibitem [{\citenamefont {Hsu}\ and\ \citenamefont {Brun}(2016)}]{Hsu_CTQEC}%
  \BibitemOpen
  \bibfield  {author} {\bibinfo {author} {\bibfnamefont {K.-C.}\ \bibnamefont
  {Hsu}}\ and\ \bibinfo {author} {\bibfnamefont {T.~A.}\ \bibnamefont {Brun}},\
  }\href@noop {} {\bibfield  {journal} {\bibinfo  {journal} {Phys. Rev. A}\
  }\textbf {\bibinfo {volume} {93}},\ \bibinfo {pages} {022321} (\bibinfo
  {year} {2016})}\BibitemShut {NoStop}%
\bibitem [{\citenamefont {Lihm}\ \emph {et~al.}(2018)\citenamefont {Lihm},
  \citenamefont {Noh},\ and\ \citenamefont {Fischer}}]{Kyungjoo_AQEC}%
  \BibitemOpen
  \bibfield  {author} {\bibinfo {author} {\bibfnamefont {J.-M.}\ \bibnamefont
  {Lihm}}, \bibinfo {author} {\bibfnamefont {K.}~\bibnamefont {Noh}}, \ and\
  \bibinfo {author} {\bibfnamefont {U.~R.}\ \bibnamefont {Fischer}},\
  }\href@noop {} {\bibfield  {journal} {\bibinfo  {journal} {Phys. Rev. A}\
  }\textbf {\bibinfo {volume} {98}},\ \bibinfo {pages} {012317} (\bibinfo
  {year} {2018})}\BibitemShut {NoStop}%
\bibitem [{\citenamefont {Touzard}\ \emph {et~al.}(2018)\citenamefont
  {Touzard}, \citenamefont {Grimm}, \citenamefont {Leghtas}, \citenamefont
  {Mundhada}, \citenamefont {Reinhold}, \citenamefont {Axline}, \citenamefont
  {Reagor}, \citenamefont {Chou}, \citenamefont {Blumoff}, \citenamefont
  {Sliwa}, \citenamefont {Shankar}, \citenamefont {Frunzio}, \citenamefont
  {Schoelkopf}, \citenamefont {Mirrahimi},\ and\ \citenamefont
  {Devoret}}]{Touzard18}%
  \BibitemOpen
  \bibfield  {author} {\bibinfo {author} {\bibfnamefont {S.}~\bibnamefont
  {Touzard}}, \bibinfo {author} {\bibfnamefont {A.}~\bibnamefont {Grimm}},
  \bibinfo {author} {\bibfnamefont {Z.}~\bibnamefont {Leghtas}}, \bibinfo
  {author} {\bibfnamefont {S.~O.}\ \bibnamefont {Mundhada}}, \bibinfo {author}
  {\bibfnamefont {P.}~\bibnamefont {Reinhold}}, \bibinfo {author}
  {\bibfnamefont {C.}~\bibnamefont {Axline}}, \bibinfo {author} {\bibfnamefont
  {M.}~\bibnamefont {Reagor}}, \bibinfo {author} {\bibfnamefont
  {K.}~\bibnamefont {Chou}}, \bibinfo {author} {\bibfnamefont {J.}~\bibnamefont
  {Blumoff}}, \bibinfo {author} {\bibfnamefont {K.~M.}\ \bibnamefont {Sliwa}},
  \bibinfo {author} {\bibfnamefont {S.}~\bibnamefont {Shankar}}, \bibinfo
  {author} {\bibfnamefont {L.}~\bibnamefont {Frunzio}}, \bibinfo {author}
  {\bibfnamefont {R.~J.}\ \bibnamefont {Schoelkopf}}, \bibinfo {author}
  {\bibfnamefont {M.}~\bibnamefont {Mirrahimi}}, \ and\ \bibinfo {author}
  {\bibfnamefont {M.~H.}\ \bibnamefont {Devoret}},\ }\href@noop {} {\bibfield
  {journal} {\bibinfo  {journal} {Physical Review X}\ }\textbf {\bibinfo
  {volume} {8}},\ \bibinfo {pages} {021005} (\bibinfo {year}
  {2018})}\BibitemShut {NoStop}%
\bibitem [{\citenamefont {Zhang}\ \emph {et~al.}(2019)\citenamefont {Zhang},
  \citenamefont {Lester}, \citenamefont {Gao}, \citenamefont {Jiang},
  \citenamefont {Schoelkopf},\ and\ \citenamefont {Girvin}}]{ZhangY19}%
  \BibitemOpen
  \bibfield  {author} {\bibinfo {author} {\bibfnamefont {Y.}~\bibnamefont
  {Zhang}}, \bibinfo {author} {\bibfnamefont {B.~J.}\ \bibnamefont {Lester}},
  \bibinfo {author} {\bibfnamefont {Y.~Y.}\ \bibnamefont {Gao}}, \bibinfo
  {author} {\bibfnamefont {L.}~\bibnamefont {Jiang}}, \bibinfo {author}
  {\bibfnamefont {R.~J.}\ \bibnamefont {Schoelkopf}}, \ and\ \bibinfo {author}
  {\bibfnamefont {S.~M.}\ \bibnamefont {Girvin}},\ }\href@noop {} {\bibfield
  {journal} {\bibinfo  {journal} {Physical Review A}\ }\textbf {\bibinfo
  {volume} {99}},\ \bibinfo {pages} {012314} (\bibinfo {year}
  {2019})}\BibitemShut {NoStop}%
\bibitem [{\citenamefont {Baumgartner}\ and\ \citenamefont
  {Narnhofer}(2008)}]{Lindbladian_dynamics}%
  \BibitemOpen
  \bibfield  {author} {\bibinfo {author} {\bibfnamefont {B.}~\bibnamefont
  {Baumgartner}}\ and\ \bibinfo {author} {\bibfnamefont {H.}~\bibnamefont
  {Narnhofer}},\ }\href@noop {} {\bibfield  {journal} {\bibinfo  {journal} {J.
  Phys. A: Math. Theor.}\ }\textbf {\bibinfo {volume} {41}},\ \bibinfo {pages}
  {395303} (\bibinfo {year} {2008})}\BibitemShut {NoStop}%
\bibitem [{\citenamefont {Albert}\ and\ \citenamefont
  {Jiang}(2014)}]{Lindbladian_dynamics_2}%
  \BibitemOpen
  \bibfield  {author} {\bibinfo {author} {\bibfnamefont {V.~V.}\ \bibnamefont
  {Albert}}\ and\ \bibinfo {author} {\bibfnamefont {L.}~\bibnamefont {Jiang}},\
  }\href@noop {} {\bibfield  {journal} {\bibinfo  {journal} {Phys. Rev. A}\
  }\textbf {\bibinfo {volume} {89}},\ \bibinfo {pages} {022118} (\bibinfo
  {year} {2014})}\BibitemShut {NoStop}%
\bibitem [{SM()}]{SM}%
  \BibitemOpen
  \href@noop {} {\ }\bibinfo {note} {See Supplemental Material}\BibitemShut
  {NoStop}%
\bibitem [{\citenamefont {Breuer}\ and\ \citenamefont
  {Petruccione}(2007)}]{Breuer_projective}%
  \BibitemOpen
  \bibfield  {author} {\bibinfo {author} {\bibfnamefont {H.-P.}\ \bibnamefont
  {Breuer}}\ and\ \bibinfo {author} {\bibfnamefont {F.}~\bibnamefont
  {Petruccione}},\ }\href@noop {} {\emph {\bibinfo {title} {The Theory of Open
  Quantum Systems}}}\ (\bibinfo  {publisher} {Oxford University Press},\
  \bibinfo {year} {2007})\BibitemShut {NoStop}%
\bibitem [{\citenamefont {Bacon}(2006)}]{Bacon_subsystem_QEC}%
  \BibitemOpen
  \bibfield  {author} {\bibinfo {author} {\bibfnamefont {D.}~\bibnamefont
  {Bacon}},\ }\href@noop {} {\bibfield  {journal} {\bibinfo  {journal} {Phys.
  Rev. A}\ }\textbf {\bibinfo {volume} {73}},\ \bibinfo {pages} {012340}
  (\bibinfo {year} {2006})}\BibitemShut {NoStop}%
\end{thebibliography}%

\onecolumngrid
\newpage
\begin{center}
	{\Large Autonomous quantum error correction and quantum computation \\ Supplementary Material}
\end{center}

				\section{Useful identities on symmetrized products} 
				Our Lemma in the main manuscript introduced symmetrized products of operators $\mathcal{S}(\{A_1,k_1\},...,\{A_n,k_n\})$. The demonstrations of our results, detailed in the subsequent sections of this supplementary, heavily rely on the manipulations of such quantities. In order to facilitate the reader we present here a few basic relations involving symmetrized products.
				\begin{itemize}
					\item Binomial expansion
					\begin{equation}
					(\sum_{i=1}^{n}A_i)^q=\sum_{k_1,...,k_n|\sum_i k_i=q}\mathcal{S}(\{A_1,k_1\},...,\{A_n,k_n\})\label{SM_eq:symmetrized_binomial}
					\end{equation}
					This relation can be seen also as a definition for symmetrized products.
					\item Invariance by permutation
					\begin{equation}
					\mathcal{S}(\{A_1,k_1\},...,\{A_n,k_n\})=\mathcal{S}(\{A_{\sigma(1)},k_{\sigma(1)}\},...,\{A_{\sigma(n)},k_{\sigma(n)}\})\label{SM_eq:symmetrized_permutation}
					\end{equation}
					for all permutation $\sigma\in S_n$ of the symmetric group $S_n$.
					\item Exponential power series expansion
					\begin{equation}
					\textrm{exp}[\sum_{i=1}^{n}A_i]=\sum_{k_1,...,k_n\in\mathbb{N}}\mathcal{S}(\{A_1,k_1\},...,\{A_n,k_n\})/(k_1+...+k_n)!\label{SM_eq:symmetrized_exponential}
					\end{equation} 
					\item Recursivity
					\begin{equation}
					\mathcal{S}(\{A_1,k_1\},...,\{A_n,k_n\})=\sum_{i=1}^{n}A_i\mathcal{S}(\{A_1,k_1\},...,\{A_i,k_i-1\},...,\{A_n,k_n\})\label{SM_eq:symmetrized_recursivity}
					\end{equation}
					In particular $\mathcal{S}(\{A_1,k_1\},...,\{A_n,k_n=0\})= \mathcal{S}(\{A_1,k_1\},...,\{A_{n-1},k_{n-1}\})$.
				\end{itemize}
				\section{Proof theorem}
				In this supplementary section we present the proof of the theorem of the main manuscript, which is formulated as follows.\\\\
				\textit{Theorem: existence of engineered dissipation and generalized error-transparent Hamiltonian.} \\
				Let us have some integer $c\geq 0$. The following statements are equivalent:
				\begin{enumerate}[label=(\arabic*)]
					\item \emph{Knill-Laflamme:} The Knill-Laflamme condition is satisfied for the error set $\mathcal{E}^{[\sim c]}$.
					\item \emph{AutoQEC:} There exists a Liouvillian  $\mathcal{L}_{\textrm{E}}$ performing autonomous quantum error correction  up to order $c$ with respect to the code space $\mathcal{C}$ and the natural dissipation $\mathcal{L}_{\textrm{n}}$.
					\item \emph{Autonomous error-corrected quantum computations:} There exists an engineered dissipation  $\mathcal{L}_{\textrm{E}}$ satisfying the following property: for all logical Hamiltonians $H_0=P_{\mathcal{C}}H_{0}P_{\mathcal{C}}$, there exists a Hamiltonian $H$ such that   $\mathcal{L}_{\textrm{E}}$  and $H$ perform an autonomous error-corrected quantum computation up to order $c$  with respect to the code space $\mathcal{C}$, the target logical Hamiltonian $H_0$, and natural dissipation $\mathcal{L}_{\rm{n}}$.
				\end{enumerate}
				
				More precisely we will prove the theorem assuming that the Lemma of the main manuscript has been proved, and postpone the proof of the Lemma to the subsequent supplementary section. First of all, the theorem hypothesis (3) trivially implies (2). 
				\subsection{$(1)\Rightarrow(3)$}
				We now assume that the theorem hypothesis (1) is true and will prove the property (3). To do so we will show  for any target Hamiltonian  $H_0$ that our explicit construction for the engineered dissipation $\mathcal{L}_{\rm{E}}$ (which is independent of $H_0$)  and the Hamiltonian $H$ in Eqs.~(\jose{8}) of the main manuscript satisfy the property (2) of the Lemma for all integers $k_1,k_2,k_3\in\mathbb{N}\times\mathbb{N}\times\{0,...,c\}$. As a preliminary result, it is helpful to note that recovery projector possesses an exact analytical expression 
				\begin{equation}
				\mathcal{P}_{\rm{E}}\rho=P_{\mathcal{C}}\rho P_{\mathcal{C}}+\sum_{n=1}^{c}\sum_{i_{n}=1}^{p_n}F_{\textrm{E},i_{n}}^{[n]}\rho F_{\textrm{E},i_{n}}^{[n]\dagger}+\sum_{q=1}^{q_{\rm{max}}}F_{\textrm{E},q}^{[\textrm{res}]}\rho F_{\textrm{E},q}^{[\textrm{res}]\dagger}\label{SM_eq:explicit_projector}
				\end{equation}
				within our construction. For some integer $n\geq0$, let us now define the set
				$$\Pi^{[\sim n]} \equiv \textrm{span}\lbrace E_{k}^{[m]} |\mu\rangle\langle\nu|(E_{k'}^{[m']})^{\dagger} | E_{k}^{[m]}\in\mathcal{E}^{[m]},E_{k'}^{[m']}\in \mathcal{E}^{[m']}, m+m'\le 2n, |\mu\rangle,|\nu\rangle \in \mathcal{C}  \rbrace.$$

				First, $\Pi^{[\sim 0]}= \mathcal{C}\otimes_d\mathcal{C}$, and  it is simple to prove that given some $\rho\in\Pi^{[\sim n]} $ one has
				\begin{eqnarray}
				\mathcal{L}_{\rm{n}}\rho&\in&\Pi^{[\sim n+1]}\label{SM_1st_id}\\
				\mathcal{L}_{\rm{E}}\rho&\in&\Pi^{[\sim n]}\\
				\mathcal{H}\rho&\in&\Pi^{[\sim n]}.
				\end{eqnarray}
				Secondly, one can prove the two additional relations 
				\begin{eqnarray}
				\mathcal{P}_{\rm{E}}\mathcal{L}_{\rm{E}}&=&0\label{SM_eq:cancellation_L_E}\\
				\mathcal{P}_{\rm{E}}\mathcal{H}&=&\mathcal{H}_0\mathcal{P}_{\rm{E}}(\mathcal{I}-\mathcal{P}_{\textrm{res}}),
				\end{eqnarray}
				where $\mathcal{P}_{\textrm{res}}\rho=P_{\textrm{res}}\rho P_{\textrm{res}}$ and $P_{\textrm{res}}=\sum_{q=1}^{q_{\rm{max}}}\ket{\phi_q}\bra{\phi_q}$ is the projector onto residual states. While the former identity is a simple consequence of the definition of the recovery projector, the latter is a consequence of the Knill-Laflamme conditions as well as of our explicit expression Eq.~(\ref{SM_eq:explicit_projector}) for the recovery projector.
				Finally, for any integer $n\leq c$ and $\rho\in \Pi^{[\sim n]}$ one has
				\begin{equation}
				(\mathcal{I}-\mathcal{P}_{\textrm{res}})\rho=\rho\label{SM_eq:identity_res}:
				\end{equation}
				In fact, to have  $\mathcal{P}_{\textrm{res}}\rho\neq 0$, $\rho$ needs to contain errors of a weight superior to $c$ on both its sides and thus has to belong in $\mathcal{H}\otimes_d\mathcal{H} \backslash\Pi^{[\sim c]} $. 
				
				In the absence of natural dissipation events ($k_3=0$),  identities Eq.~(\ref{SM_1st_id}-\ref{SM_eq:identity_res}) yield 
				\begin{equation}
				\mathcal{P}_{\rm{E}}\mathcal{S}[\{\mathcal{H},k_1\},\{\mathcal{L}_{\rm{E}},k_2\},\{\mathcal{L}_{\rm{n}},0\}]\mathcal{P}_{\mathcal{C}}= \delta_{k_2,0}\mathcal{H}_0^{k_1}\mathcal{P}_{\rm{E}}\mathcal{P}_{\mathcal{C}}=\delta_{k_2,0}\mathcal{H}_0^{k_1}\mathcal{P}_{\mathcal{C}}
				\end{equation}
				which is the desired result. In the presence of natural dissipation events ($k_3\neq 0$), the following recursive relation connects  $k_3$-th order of natural dissipation error to $(k_3-1)$-th order  for any $\rho\in\mathcal{C}\otimes_d\mathcal{C}$: 
				\begin{equation}\mathcal{P}_{\rm{E}}\mathcal{S}[\{\mathcal{H},k_1\},\{\mathcal{L}_{\rm{E}},k_2\},\{\mathcal{L}_{\rm{n}},k_3\}]\rho=\sum_{q=0}^{k_1}\mathcal{H}_{0}^{q}\mathcal{P}_{\rm{E}} \mathcal{L}_{\rm{n}}\rho^{(q)}\label{SM_eq:higher-order_processes},
				\end{equation}
				where $\rho^{(q)}=\mathcal{S}[\{\mathcal{H},k_1-q\},\{\mathcal{L}_{\rm{E}},k_2\},\{\mathcal{L}_{\rm{n}},k_3-1\}]\rho\in  \Pi^{[\sim k_3-1]}\subset\Pi^{[\sim c-1]}$. Finally, exploiting the Knill-Laflamme conditions and our explicit construction one proves below that 
				\begin{equation}
				\mathcal{P}_{\rm{E}} \mathcal{L}_{\rm{n}}\rho=0\qquad\forall\rho \in \Pi^{[\sim c-1]}\label{SM_eq:key_identity}
				\end{equation}
				which combined to Eq.~(\ref{SM_eq:higher-order_processes}) leads to the property (2) of the Lemma. 
				
				We will now complete the proof by demonstrating the validity  of the key identity Eq.~(\ref{SM_eq:key_identity}). We will prove this for any matrix of the form $\rho=E_k^{[m]}\ket{\mu_\alpha}\bra{\mu_\beta}E_{k'}^{[m']\dagger}$ for any integers $m,m'$ such that $m+m'\leq  2(c-1)$, $E_k^{[m]},E_{k'}^{[m']}$ respectively in $\mathcal{E}^{[m]}$ and $\mathcal{E}^{[m']}$ and $\alpha,\beta\in\{1,...,d_{\mathcal{C}}\}$. The desired result will then proceed simply by linearity. Since, $\mathcal{L}_{\rm{n}}(\rho)\in\Pi^{[\sim c]}$, then $F_{\textrm{E},q}^{[\textrm{res}]}(\mathcal{L}_{\rm{n}}\rho)F_{\textrm{E},q}^{[\textrm{res}]\dagger}=0$ for all $q=1,...,q_{\textrm{max}}$, and one gets
				\begin{eqnarray}
				\mathcal{P}_{\rm{E}} \mathcal{L}_{\rm{n}}\rho&=&P_{\mathcal{C}}(\mathcal{L}_{\rm{n}}\rho) P_{\mathcal{C}}+\sum_{n=1}^{c}\sum_{i_{n}=1}^{p_n}F_{\textrm{E},i_{n}}^{[n]}(\mathcal{L}_{\rm{n}}\rho) F_{\textrm{E},i_{n}}^{[n]\dagger}\nonumber\\
				&=&\sum_{a,b=1}^{d_{\mathcal{C}}}\sum_{n=0}^{c}\sum_{i_{n}=0}^{p_{n}}|\mu_{a}\rangle\langle\mu_{b}| \langle \mu_{a,i_{n}}^{[n]}| (\mathcal{L}_{\rm{n}}\rho) |\mu^{[m]}_{b,i_{m}}\rangle\nonumber\\
				&=&\sum_{l=1}^{N}\sum_{a,b=1}^{d_{\mathcal{C}}}\sum_{n=0}^{c}\sum_{i_{n}=0}^{p_{n}}|\mu_{a}\rangle\langle\mu_{b}| \times{\Big \lbrace} \langle\mu_{a,i_{n}}^{[n]}|F_l E_k^{[m]}|\mu_\alpha\rangle  \langle\mu_\beta|E_{k'}^{[m']\dagger}  F_l ^\dagger|\mu_{b,i_{n}}^{[n]}\rangle \nonumber\\
				&&-\frac{1}{2}\left[\langle\mu_{a,i_{n}}^{[n]}|F_l ^\dagger F_l E_k^{[m]}|\mu_\alpha\rangle  \langle\mu_\beta|E_{k'}^{[m']\dagger}|\mu_{b,i_{n}}^{[n]}\rangle+\langle\mu_{a,i_{n}}^{[n]}|E_k^{[m]}|\mu_\alpha\rangle  \langle\mu_\beta|E_{k'}^{[m']\dagger}F_l ^\dagger F_l |\mu_{b,i_{n}}^{[n]}\rangle\right]{\Big \rbrace}\label{SM_eq:calculation_projector_pi}
				\end{eqnarray} 
				Moreover, as a consequence of the Knill-Laflamme conditions, one gets that the orthonormalized error states generated by the Gramm-Schmidt algorithm  in the main manuscript satisfy
				\begin{equation}
				|\mu_{a,i_n}^{[n]}\rangle=\sum_{E\in\mathcal{E}^{[n]}}\Gamma_{E,i_n}^{[n]}E\ket{\mu_a}\label{SM_eq:consequence_KL}
				\end{equation}
				as long as $n\leq c$, where $\Gamma_{E,i_n}^{[n]}\in\mathbb{C}$ is a constant independent of the code space basis state $\ket{\mu_a}$  ($a\in\{1,...,d_{\mathcal{C}}\}$). From this one deduces further for any $E\in\mathcal{E}^{[\sim c]}$ and $n\leq c$ that $\langle \mu_{a,i_n}^{[n]}|E|\mu_b\rangle=\Omega_{i_n,n}^{E}\delta_{a,b}$ with $\Omega^E_{i_n,n}=\langle \mu_{a,i_n}^{[n]}|E|\mu_a\rangle$ is independent of $a$. We now assume that $m<m'$ (the demonstration in the cases $m=m'$ and $m>m'$ is very similar) and thus in particular $m\leq c-2$: thus $E_k^{[m]}$, $F_l E_k^{[m]}$ as well as $F_l^{\dagger}F_l E_k^{[m]}$ all belong in $\mathcal{E}^{[\sim c]}$. One can insert the latter result in Eq.~(\ref{SM_eq:calculation_projector_pi}), yielding:
				\begin{eqnarray}
				\mathcal{P}_{\rm{E}} \mathcal{L}_{\rm{n}}\rho&=&\sum_{l=1}^{N}\sum_{a=1}^{d_{\mathcal{C}}}\sum_{b=1}^{d_{\mathcal{C}}}\sum_{n=0}^{c}\sum_{i_{n}=0}^{p_{n}}\delta_{a,\alpha}|\mu_{\alpha}\rangle\langle\mu_{b}| \times{\Big \lbrace} \langle\mu_{\alpha,i_{n}}^{[n]}|F_l E_k^{[m]}|\mu_\alpha\rangle  \langle\mu_\beta|E_{k'}^{[m']\dagger}  F_l ^\dagger|\mu_{b,i_{n}}^{[n]}\rangle \nonumber\\
				&&-\frac{1}{2}\left[\langle\mu_{\alpha,i_{n}}^{[n]}|F_l ^\dagger F_l E_k^{[m]}|\mu_\alpha\rangle  \langle\mu_\beta|E_{k'}^{[m']\dagger}|\mu_{b,i_{n}}^{[n]}\rangle+\langle\mu_{\alpha,i_{n}}^{[n]}|E_k^{[m]}|\mu_\alpha\rangle  \langle\mu_\beta|E_{k'}^{[m']\dagger}F_l ^\dagger F_l |\mu_{b,i_{n}}^{[n]}\rangle\right]{\Big \rbrace}\nonumber\\
				&=&\sum_{l=1}^{N}\sum_{b=1}^{d_{\mathcal{C}}}\sum_{n=0}^{c}\sum_{i_{n}=0}^{p_{n}}|\mu_{\alpha}\rangle\langle\mu_{b}| \times{\Big \lbrace} \langle\mu_{b,i_{n}}^{[n]}|F_l E_k^{[m]}|\mu_b\rangle  \langle\mu_\beta|E_{k'}^{[m']\dagger}  F_l ^\dagger|\mu_{b,i_{n}}^{[n]}\rangle \nonumber\\
				&&-\frac{1}{2}\left[\langle\mu_{b,i_{n}}^{[n]}|F_l ^\dagger F_l E_k^{[m]}|\mu_b\rangle  \langle\mu_\beta|E_{k'}^{[m']\dagger}|\mu_{b,i_{n}}^{[n]}\rangle+\langle\mu_{b,i_{n}}^{[n]}|E_k^{[m]}|\mu_b\rangle  \langle\mu_\beta|E_{k'}^{[m']\dagger}F_l ^\dagger F_l |\mu_{b,i_{n}}^{[n]}\rangle\right]{\Big \rbrace}\nonumber\\
				&=&\sum_{l=1}^{N}\sum_{b=1}^{d_{\mathcal{C}}}|\mu_{\alpha}\rangle\langle\mu_{b}| \times{\Big \lbrace}   \langle\mu_\beta|E_{k'}^{[m']\dagger}  F_l ^\dagger\sum_{n=0}^{c}\sum_{i_{n}=0}^{p_{n}}\left[|\mu_{b,i_{n}}^{[n]}\rangle \langle\mu_{b,i_{n}}^{[n]}|\right]F_l E_k^{[m]}|\mu_b\rangle\nonumber\\
				&&-\frac{1}{2}\left[ \langle\mu_\beta|E_{k'}^{[m']\dagger}\sum_{n=0}^{c}\sum_{i_{n}=0}^{p_{n}}\left[|\mu_{b,i_{n}}^{[n]}\rangle\langle\mu_{b,i_{n}}^{[n]}|\right]F_l ^\dagger F_l E_k^{[m]}|\mu_b\rangle + \langle\mu_\beta|E_{k'}^{[m']\dagger}F_l ^\dagger F_l \sum_{n=0}^{c}\sum_{i_{n}=0}^{p_{n}}\left[|\mu_{b,i_{n}}^{[n]}\rangle \langle\mu_{b,i_{n}}^{[n]}|\right]E_k^{[m]}|\mu_b\rangle \right]{\Big \rbrace}\nonumber\\
				&=&\sum_{l=1}^{N}\sum_{b=1}^{d_{\mathcal{C}}}|\mu_{\alpha}\rangle\langle\mu_{b}| \times{\Big \lbrace}   \langle\mu_\beta|E_{k'}^{[m']\dagger}  F_l ^\dagger F_l E_k^{[m]}|\mu_b\rangle-\frac{1}{2}\left[ \langle\mu_\beta|E_{k'}^{[m']\dagger}F_l ^\dagger F_l E_k^{[m]}|\mu_b\rangle + \langle\mu_\beta|E_{k'}^{[m']\dagger}F_l ^\dagger F_l E_k^{[m]}|\mu_b\rangle \right]{\Big \rbrace}\nonumber\\
				&=&0
				\end{eqnarray}
				where we used the fact that $\sum_{n=0}^{c}\sum_{i_{n}=0}^{p_{n}}|\mu_{b,i_{n}}^{[n]}\rangle \langle\mu_{b,i_{n}}^{[n]}|E|\mu_{b}\rangle=E |\mu_b\rangle$ for any $E \in\mathcal{E}^{[\sim c]}$.\hfill$\square$
				\subsection{$(2)\Rightarrow(1)$}
				We now assume (2) and will prove that Knill-Laflamme conditions are satisfied for the error set $\mathcal{E}^{[c]}$. Thus there exists an engineered dissipation $\mathcal{L}_{\rm{E}}$ performing AutoQEC up to order $c$. In particular, the statement (2) of the Lemma is satisfied for $H_0=H=0$, which for $k_1=k_2=0$ yields $\mathcal{P}_{\rm{E}}\mathcal{L}_{\rm{n}}^k\mathcal{P}_{\mathcal{C}}=\delta_{k,0}\mathcal{P}_{\mathcal{C}}$ for all $k\in\{0,...,c\}$. We will show from this that $\mathcal{P}_{\rm{E}}$ is a good quantum recovery in the traditional sense of non-autonomous QEC for the error set $\mathcal{E}^{[c]}$, a result known to be enforce the Knill-Laflamme condtions to be satsfied onto that same error set. First $\mathcal{L}_{\rm{E}}$ being a Lindbladian, the recovery projector $\mathcal{P}_{\rm{E}}=\textrm{lim}_{t\to+\infty}[\mathcal{L}_{\rm{E}}t]$ is a CPTP map and it admits a Kraus representation: $\mathcal{P}_{\rm{E}}=\sum_{l=1}^{L} R_l \bullet R_l^\dagger$ with $\sum_{l=1}^{L} R_l^\dagger R_l =\mathbb{1}$. We then prove inductively over $k\in\{0,...,c\}$ the following statement:	
				\begin{center}
					\textit{For all $E\in \mathcal{E}^{[\sim k]}$ and $l\in\{1,L\}$,  one has $R_l E P_{\mathcal{C}}=\Gamma_{l,E}P_{\mathcal{C}}$ for some constant $\Gamma_{l,E}$.} 
				\end{center}
				First, one has  $P_{\mathcal{C}}\bullet P_{\mathcal{C}}=\mathcal{P}_{\rm{E}}\mathcal{P}_{\mathcal{C}}=\sum_{l=1}^{L} R_l P_{\mathcal{C}} \bullet P_{\mathcal{C}}R_l^\dagger$. By unicity of the Kraus operators of quantum channels up to unitaries, we deduce that  $\forall l\in\{1,L\}$,  exists $\Gamma_{l,\mathcal{D}_{\mathbb{1}}}$ such that $R_l\mathbb{1}P_{\mathcal{C}}=\Gamma_{l,\mathbb{1}}P_{\mathcal{C}}$. Since the identity is the only error in $\mathcal{E}^{[\sim k=0]}$ one gets that the desired statement is true for $k=0$. One now assume that the statement is true up to $k-1\leq c-1$. Expanding natural dissipation to order  $k$ yields:
				\begin{equation}
				\mathcal{L}_{\textrm{n}}^k=\sum_{k'=0}^{k}\left(\frac{-1}{2}\right)^{k-k'}\hspace{-20pt}\sum_{(d_i)\in \{1,..,N\}^{k'}}\underset{\sum_i n_i+m_i=k-k'}{\sum_{(n_i),(m_i)\in \mathbb{N}^{k'+1}}}\mathcal{B}_{(n_i),(d_i)}
				\end{equation}
				where $\mathcal{B}_{(n_i),(d_i)}=\prod_{i=1}^{k'+1}[C_{n_i+m_i}^{n_i}]E_{(d_i),(n_i)}\bullet E_{(d_i),(m_i)}^{\dagger}$, $C_{n}^{k}$ is the binomial coefficient
				and $E_{(d_i),(n_i)}\equiv H_{\textrm{n,BA}}^{n_{k'+1}}\prod_{i=1}^{k'}(F_{d_i}H_{\textrm{n,BA}}^{n_{i}})$. Let us have $\ket{\mu}\in\mathcal{C}$, and $\ket{\nu}\in\mathcal{H}$ such that  $\ket{\nu}\perp\ket{\mu}$. Considering the identity $0=\bra{\nu}(\mathcal{P}_{\rm{E}}\mathcal{L}_{\textrm{n}}^k[\ket{\mu}\bra{\mu}])|\ket{\nu}$, and applying the inductive hypothesis for $k-1$ one gets:
				
				\begin{eqnarray}
				0&=&\sum_{l=1}^{L}\sum_{{\tiny \begin{array}{c}k'=0\\k-k'\text{ even}\end{array}}}^{k}\frac{1}{2^{k-k'}}\sum_{\underset{d_i\in \{1,...,N\}}{(d_i)_{1\leq i\leq k'}}}\sum_{{\tiny \begin{array}{c}(n_i,m_i)_{1\leq i\leq k'+1}\\ \sum_i n_i=\sum_i m_i=\frac{k-k'}{2}\end{array}}}\left[\prod_{i=1}^{k'+1}C_{n_i+m_i}^{n_i}\right]\bra{\nu}R_l E_{(d_i),(n_i)}P_{\mathcal{C}}\ket{\mu}\bra{\mu} P_{\mathcal{C}}E_{(d_i),(m_i)}^{\dagger}R_l^\dagger\ket{\nu}\label{eq:symmetric_KL}
				\end{eqnarray}
				Note that the scalar products in Eq.~(\ref{eq:symmetric_KL}) only involves errors $E_{(d_i),(n_i)}$  and $E_{(d_i),(m_i)}$ of identical weight $k$, where the backaction Hamiltonian $H_{\textrm{n,BA}}$ has been applied the same amount of times $(k-k')/2$ on both left and right sides, and  $k'$ jump operators $F_{d_i}$ have  been applied simultaneously on both sides: indeed since $\ket{\mu}\in\mathcal{C}$ and $\ket{\nu}\perp\ket{\mu}$ one can show via the inductive hyptothesis that assymetric terms with unequal weight must vanish (as either $\bra{\nu}R_l E_{(d_i),(n_i)}P_{\mathcal{C}}\ket{\mu}$ or $\bra{\mu} P_{\mathcal{C}}E_{(d_i),(m_i)}^{\dagger}R_l^\dagger\ket{\nu}$ necessarily  involve  an error of weight inferior or equal to $k-1$). Eq.~(\ref{eq:symmetric_KL}) can be reformulated in the more compact form
				\begin{equation}
				0=\sum_{l=1}^{L}\sum_{{\tiny \begin{array}{c}k'=0\\k-k'\textrm{ even}\end{array}}}^k\left(\frac{1}{2}\right)^{k-k'}\hspace{-20pt}\sum_{(d_i)\in \{1,..,N\}^{k'}}\bra{X_{l,(d_i)}}\mathcal{M}_k\ket{X_{l,(d_i)}}\label{SM_eq:induction}
				\end{equation}
				where $\mathcal{M}_k=\bigotimes_{i=1}^{k'+1}\mathcal{M}_k^{(i)}$,  $\mathcal{M}_k^{(i)}=\sum_{n,m=0}^{(k-k')/2}C_{n+m}^{n}\ket{n}\bra{m}$  is the  symmetric Pascal matrix defined over some abstract quantum system $\tilde{\mathcal{H}}_k=\textrm{span}[|n\rangle,0\leq n\leq k-k'/2]$ of dimension $(k-k')/2+1$ and
				$$\ket{X_{l,(d_i),(n_i)}}=\sum_{(n_i)\in \mathbb{N}^{k'+1}| \sum_i n_i=(k-k')/2} \bra{\nu}R_l E_{(d_i),(n_i)}P_{\mathcal{C}}\ket{\mu} \bigotimes_{i=1}^{k'+1}\ket{n_i}$$
				is a vector representing a state in a tensor product of $k'+1$ copies of such system. One notices that $\mathcal{M}^{(i)}=\textrm{exp}(A)^\dagger \textrm{exp}(A)$, where $A=\sum_{n=1}^{(k-k')/2}n\ket{n-1}\bra{n}$, and thus the Pascal matrix is Hermitian positive definite. 
				From this, one concludes that $\bra{\nu}R_l E P_{\mathcal{C}}\ket{\mu}=0$ for all $l\in\{1,...,L\}$ and $E\in \mathcal{E}^{[k]}$ (indeed all the errors in $\mathcal{E}^{[k]}$ can be written as  $E_{(d_i),(n_i)}$ for some sequences $(d_i)_{1\leq i\leq k'}$ and $(n_i)_{1\leq i\leq k'+1}$). The previous result being true for all $\ket{\nu}\in\mathcal{H}$ such that $\ket{\mu}\perp\ket{\nu}$, one deduces that 
				\begin{equation}
				R_l E\ket{\mu}=\gamma_{\mu}\ket{\mu}.\label{SM_eq:proportionality_result}
				\end{equation}
				Since  Eq.~(\ref{SM_eq:proportionality_result})  is true  for all $\ket{\mu}\in\mathcal{C}$ the proportionality constant $\gamma_{\mu}$ needs to be independent of the specific code state $\ket{\mu}$. One concludes that there exists $\Gamma_{l,E}$ such that $R_lEP_{\mathcal{C}}=\Gamma_{l,E}P_{\mathcal{C}}$. Thus one has proved that the desired inductive statement is also true for $k$, and thus is true up to $k=c$. As a conclusion one deduces for all $E,E'\in \mathcal{E}^{[\sim c]}$ that
				$P_{\mathcal{C}}E'^\dagger EP_{\mathcal{C}}=P_{\mathcal{C}}E'^\dagger(\sum_{l=1}^{L}R_l^\dagger R_l )E =\Gamma_{E',E}P_{\mathcal{C}},$
				where $\Gamma_{\mathcal{E}',\mathcal{E}}=\sum_l \Gamma_{l,\mathcal{E}'}^*\Gamma_{l,\mathcal{E}}$ which is precisely the Knill-Laflamme condition.\hfill $\square$
				\section{Proof Lemma}
				In this supplementary section we present  the proof of the Lemma of the main manuscript, which is formulated as follows.\\\\
				\textit{Lemma: Properties of generalized error-transparent Hamiltonians and engineered dissipation.} Let us have some target Hamiltonian $H_{0}=P_{\mathcal{C}}H_{0}P_{\mathcal{C}}$, some engineered dissipation $\mathcal{L}_{\rm{E}}$ and generic Hamiltonian $H$.
				We denote $\mathcal{H}$ (resp. $\mathcal{H}_0$ ) as the superoperator associated with the Hamiltonian evolution  $\mathcal{H}(\rho)=[H,\rho]$ (resp. $\mathcal{H}_0(\rho)=[H_0,\rho]$). The following statements are equivalent:
				\begin{enumerate}[label=(\arabic*)]
					\item  $\mathcal{L}_{\textrm{E}}$  and $H$ perform an autonomous error-corrected quantum computation up to order $c$  with respect to the code space $\mathcal{C}$, the target logical Hamiltonian $H_0$ and natural dissipation $\mathcal{L}_{\rm{n}}$.
					\item for all sets of integers $k_1,k_2,k_3\in\mathbb{N}\times\mathbb{N}\times\{0,...,c\}$ one has 
					\begin{equation}
					\mathcal{P}_{\rm{E}}\mathcal{S}[\{\mathcal{H},k_1\},\{\mathcal{L}_{\rm{E}},k_2\},\{\mathcal{L}_{\rm{n}},k_3\}]\mathcal{P}_{\mathcal{C}}=
					\delta_{k_2,0}\delta_{k_3,0}\mathcal{H}_0^{k_1}\mathcal{P}_{\mathcal{C}}\label{eq:lemma}
					\end{equation}
				\end{enumerate}
				\subsection{$(1)\Rightarrow(2)$}
				We now assume that $(1)$ is true and thus $\mathcal{L}_{\rm{E}}$ and $H$ perform an autonomous error-corrected quantum computation up to order $c$ wrt the Hamiltonian $H_0$. Let us have $M>0$ and $g_0>0$ the positive constants involved in the definition of the latter property (see main manuscript). Let us have some dimensionless numbers $u,\tilde{g}>0$ such that $|\tilde{g}|< g_0$. For all $R>0$, we  will consider the time $t=u/(R\kappa)$  and  the Hamiltonian coupling strength $g=R\tilde{g}$. For all $\rho(0)$   in the code space and for all $R>0$ one has
				\begin{eqnarray}
				\left\|\mathcal{P}_{\rm{E}}\text{exp}[(\mathcal{L}_{\rm{E}}+\mathcal{L}_{\rm{n}}/R-i\tilde{g}\mathcal{H})u]\rho(0)-\text{exp}[-i\tilde{g}\mathcal{H}_0 u]\rho(0)\right\|&=&\left\|\mathcal{P}_{\rm{E}}\rho(t)-\text{exp}[-ig\kappa\mathcal{H}_0 t]\rho(0)\right\|\\
				&\leq& M\frac{u}{R^{c+1}}\|\rho(0)\|
				\end{eqnarray}
				which in the limit $R\to+\infty$ gives:
				$\left\|\mathcal{P}_{\rm{E}}\text{exp}[(\mathcal{L}_{\rm{E}}-i\tilde{g}\mathcal{H})u]\rho(0)-\text{exp}[-i\tilde{g}\mathcal{H}_0 u]\rho(0)\right\|\leq0$ and thus
				$\mathcal{P}_{\rm{E}}\text{exp}[(\mathcal{L}_{\rm{E}}-i\tilde{g}\mathcal{H})u]\rho(0)=\text{exp}[-i\tilde{g}\mathcal{H}_0 u]\rho(0)$. This property holds for all initial  conditions in the code space, as  well as all  dimensionless $u$ and $\tilde{g}$ such that $|\tilde{g}|\leq g_0$ enabling thus an identification between the  left and right handsides of all terms with identical powers in $g$ and $\kappa$, which according to the relations Eqs.~(\ref{SM_eq:symmetrized_exponential},\ref{SM_eq:symmetrized_recursivity}) on symmetrized products yields
				\begin{equation}
				\mathcal{P}_{\rm{E}}\mathcal{S}(\{\mathcal{H},k_1\},\{\mathcal{L}_{\rm{E}},k_2\},\{\mathcal{L}_{\rm{n}},k_3=0\})\mathcal{P}_{\mathcal{C}}=\mathcal{P}_{\rm{E}}\mathcal{S}(\{\mathcal{H},k_1\},\{\mathcal{L}_{\rm{E}},k_2\})\mathcal{P}_{\mathcal{C}}=\delta_{k_2,0}\mathcal{H}_0^{k_1}\mathcal{P}_{\mathcal{C}}\label{SM_eq:preliminary_error_transparent}
				\end{equation}
				for all $k_1,k_2\in\mathbb{N}\times\mathbb{N}$ which is the desired result for $k_3=0$. Reasoning by reductio ad absurdum, we now will make the hypothesis that condition $(2)$ of the Lemma is not true and will reach ultimately a logical  contradiction. Therefore,  there exists $k_1,k_2,k_3\in\mathbb{N}\times\mathbb{N}\times\{0,...,c\}$ such that
				\begin{equation}
				\mathcal{P}_{\rm{E}}\mathcal{S}(\{\mathcal{H},k_1\},\{\mathcal{L}_{\rm{E}},k_2\},\{\mathcal{L}_{\rm{n}},k_3\})\mathcal{P}_{\mathcal{C}}\neq\delta_{k_2,0}\delta_{k_3,0}\mathcal{H}_0^{k_1}\mathcal{P}_{\mathcal{C}}
				\end{equation}
				Due to Eq.~(\ref{SM_eq:preliminary_error_transparent}), this implies $k_3>0$. One defines $k_3^M\leq c$  as the minimum of the non-empty set: 
				\begin{equation}
				\mathcal{A}=\left\{k_3\in\{1,...,c\}| \exists (k_1,k_2)\in\mathbb{N}^2\text{ satisfying: }\mathcal{S}(\{\mathcal{H},k_1\},\{\mathcal{L}_{\rm{E}},k_2\},\{\mathcal{L}_{\rm{n}},k_3\})\mathcal{P}_{\mathcal{C}}\neq 0\right\}.
				\end{equation} 
				Based on Eq.~(\ref{SM_eq:preliminary_error_transparent}), one gets by expanding  $\text{exp}[(\mathcal{L}_{\rm{E}}+\mathcal{L}_{\rm{n}}/R-i\tilde{g}\mathcal{H})u]$  in symmetrized products
				\begin{eqnarray}
				\mathcal{P}_{\rm{E}}\rho(t)&=&\text{exp}[-ig\mathcal{H}_0\kappa t]\rho(0)+\left(\frac{u}{R}\right)^{k_3^M}\sum_{k_1,k_2}\frac{\mathcal{P}_{\rm{E}}\mathcal{S}(\{\mathcal{H},k_1\},\{\mathcal{L}_{\rm{E}},k_2\},\{\mathcal{L}_{\rm{n}},k_3^M\})\rho(0)}{(k_1+k_2+k_3^M)!}(-i \tilde{g})^{k_1} u^{k_1+k_2}\nonumber\\
				&&+\left(\frac{u}{R}\right)^{k_3^M+1}\sum_{k_1,k_2,k_3}\frac{\mathcal{P}_{\rm{E}}\mathcal{S}(\{\mathcal{H},k_1\},\{\mathcal{L}_{\rm{E}},k_2\},\{\mathcal{L}_{\rm{n}},k_3^M+1+k_3\})\rho(0)}{(k_1+k_2+k_3^M+1+k_3)!} (-i \tilde{g})^{k_1} u^{k_1+k_2+k_3}\left(\frac{1}{R}\right)^{k_3}\nonumber\\
				\end{eqnarray}
				Given that $k_3^M\in\mathcal{A}$ there exists $u_1,\tilde{g}_1> 0$ with $\tilde{g}_1<g_0$ such that $\Delta=\sum_{k_1,k_2}\frac{\mathcal{P}_{\rm{E}}\mathcal{S}(\{\mathcal{H},k_1\},\{\mathcal{L}_{\rm{E}},k_2\},\{\mathcal{L}_{\rm{n}},k_3^M\})\rho(0)}{(k_1+k_2+k_3^M)!}(-i \tilde{g}_1)^{k_1}u_1^{k_1+k_2}\neq 0$. Thus, $\forall R>0$,  defining $t=u_1/(R\kappa)$ and $g=R\tilde{g}_1<g_0 R$, one has:
				\begin{eqnarray}
				\mathcal{P}_{\rm{E}}\rho(t)-\text{exp}[-ig\mathcal{H}_0\kappa t]\rho(0)&=&\left(\frac{u_1}{R}\right)^{k_3^M}\Delta+\left(\frac{u_1}{R}\right)^{k_3^M+1}\tilde{\Delta}(R)\\
				&=&\frac{\kappa t}{R^{k_3^M-1}}u_1^{k_3^M-1}\Delta+\frac{\kappa t}{R^{k_3^M}} u_1^{k_3^M}\tilde{\Delta}(R).
				\end{eqnarray}
				with
				\begin{eqnarray}
				\Delta&=&\sum_{k_1,k_2}\frac{\mathcal{P}_{\rm{E}}\mathcal{S}(\{\mathcal{H},k_1\},\{\mathcal{L}_{\rm{E}},k_2\},\{\mathcal{L}_{\rm{n}},k_3^M\})\rho(0)}{(k_1+k_2+k_3^M)!}(-i \tilde{g}_1)^{k_1} u_1^{k_1+k_2}\neq 0\\
				\tilde{\Delta}(R)&=&\sum_{k_1,k_2,k_3}\frac{\mathcal{P}_{\rm{E}}\mathcal{S}(\{\mathcal{H},k_1\},\{\mathcal{L}_{\rm{E}},k_2\},\{\mathcal{L}_{\rm{n}},k_3^M+1+k_3\})\rho(0)}{(k_1+k_2+k_3^M+1+k_3)!} (-i \tilde{g})^{k_1} u_1^{k_1+k_2+k_3}\left(\frac{1}{R}\right)^{k_3}
				\end{eqnarray}
				Let us have some $R_0>0$. A careful analysis shows that $\tilde{\Delta}(R)$ is necessarily bounded in the domain $[R_0,+\infty[$. One thus finds that the projected density matrix verifies the asymptotic behavior:
				\begin{equation}
				\mathcal{P}_{\rm{E}}\rho(t=u_1/(R\kappa))-\rho(0)\underset{R\to+\infty}{\sim}\frac{\kappa t}{R^{k_3^M-1}}u_1^{k_3^M-1}\Delta
				\end{equation}
				Since $k_3^M-1< c$ , one concludes that for all $M>0$ there exists a  specific choice  of  time $t=u_1/(R\kappa)$ and  couplings $g=R\tilde{g}_1$, $R\geq0$ satisfying
				$\|\mathcal{P}_{\rm{E}}\rho(t)-\text{exp}[-ig\mathcal{H}_0\kappa t]\rho(0)\|>M\kappa t /R^{c}$, which is in logical contradiction with our initial  hypothesis. One  concludes that $\mathcal{S}(\{\mathcal{H},k_1\},\{\mathcal{L}_{\rm{E}},k_2\},\{\mathcal{L}_{\rm{n}},k_3\})\mathcal{P}_{\mathcal{C}}=\delta_{k_2,0}\delta_{k_3,0}\mathcal{H}_0^{k_1}\mathcal{P}_{\mathcal{C}}$.\hfill $\square$
				\subsection{$(2)\Rightarrow(1)$ }\label{sec:2_1_lemma}
				We will make use of the following identity
				\begin{equation}
				e^{(A+B)t}=e^{At}+\int_0^t  ds  e^{(A+B)(t-s)}Be^{As}.
				\end{equation}
				Given some integer $N>1$ integer,  this can be recursively generalized as:
				\begin{equation}
				e^{(A+B)t}=e^{At}+\sum_{k=1}^{N-1}\underset{\sum_i \tau_i\leq t}{\iint\left(\prod_{i=1}^{k}d\tau_{i}\right)}e^{A(t-\sum_i\tau_i)}\prod_{i=1}^{k}\left(Be^{A\tau_i}\right)
				+\underset{\sum_i \tau_i\leq t}{\iint\left(\prod_{i=1}^{N}d\tau_{i}\right)}e^{[A+B](t-\sum_i\tau_i)}\prod_{i=1}^{N}\left(Be^{A\tau_i}\right).\label{SM_eq:identity_non-perturbative}
				\end{equation}
				We denote $\mathcal{G}(t)=e^{\kappa(R\mathcal{L}_{\rm{E}}+\mathcal{L}_{\rm{n}}-ig \mathcal{H})t}$, $\mathcal{G}_1(t)=e^{\kappa( R\mathcal{L}_{\rm{E}}-ig  \mathcal{H})t}$ and $\mathcal{G}_0(t)=e^{c\kappa\mathcal{L}_{\rm{E}}t}$. Using identity Eq.~(\ref{SM_eq:identity_non-perturbative}) we get:
				\begin{eqnarray}
				\mathcal{P}_{\rm{E}}\rho(t)&=&\mathcal{P}_{\rm{E}}\mathcal{G}(t)\rho(0)\nonumber\\
				&=&\underset{=\mathcal{P}_{\rm{E}}\mathcal{G}_1(t)\mathcal{P}_{\mathcal{C}}\rho(0)}{\underbrace{\mathcal{P}_{\rm{E}}\mathcal{G}_1(t)\rho(0)}}+\sum_{k=1}^{c}\kappa^k\underset{\sum_i \tau_i\leq t}{\iint\left(\prod_{i=1}^{k}d\tau_{i}\right)}\mathcal{P}_{\rm{E}}\mathcal{G}_1\left(t-{\tiny \sum}_i\tau_i\right)\prod_{i=1}^{k}\left(\mathcal{L}_{\rm{n}}\mathcal{G}_1(\tau_i)\right)\rho(0)\nonumber\\
				&&+\kappa^{c+1}\underset{\sum_i \tau_i\leq t}{\iint\left(\prod_{i=1}^{c+1}d\tau_{i}\right)}\mathcal{P}_{\rm{E}}\mathcal{G}\left(t-{\tiny \sum}_i\tau_i\right)\prod_{i=1}^{c+1}\left(\mathcal{L}_{\rm{n}}\mathcal{G}_1(\tau_i)\right)\rho(0).\label{SM_eq:non-perturbative_error-corrected}
				\end{eqnarray}
				
				Having assumed the hypothesis (2) of the lemma to  be satisfied, from Eq.~(\ref{SM_eq:symmetrized_exponential}) we deduce $\mathcal{P}_{\rm{E}}\mathcal{G}_1(t)\mathcal{P}_{\mathcal{C}}=e^{-i\mathcal{H}_0 t}\mathcal{P}_{\mathcal{C}}$. Then, expanding $\mathcal{G}_1(t)$ in powers of the Hamiltonian superoperator $ \mathcal{H}$ as following
				\begin{equation}
				\mathcal{G}_1(t)=\mathcal{G}_0(t)+\sum_{q=1}^{+\infty}(-i\kappa g)^q\underset{\sum_i \tau_i\leq t}{\iint\left(\prod_{i=1}^{q}d\tau_{i}\right)}\mathcal{G}_0\left(t-{\tiny \sum}_i\tau_i\right)\prod_{i=1}^{q}\left(\mathcal{H}\mathcal{G}_0(\tau_i)\right)
				\end{equation}
				we get 
				\begin{multline}
					\mathcal{P}_{\rm{E}}\rho(t)-e^{-i\mathcal{H}_0 t}\rho(0)=\sum_{k=1}^{c}\sum_{q=0}^{+\infty}\kappa^k(-i\kappa g)^q\sum_{(\mathcal{A}_i)_{1\leq n\leq k+q}\in\mathcal{ O}_{k,q}}\underset{\sum_i \tilde{\tau}_i\leq t}{\iint\left(\prod_{i=1}^{k+q}d\tilde{\tau}_{i}\right)}\mathcal{P}_{\rm{E}}\mathcal{G}_0\left(t-{\tiny \sum}_i\tilde{\tau}_i\right)\prod_{i=1}^{k+q}\left(\mathcal{A}_{i}\mathcal{G}_0(\tilde{\tau}_i)\right)\rho(0)\\
					+\kappa^{c+1}\sum_{q=0}^{+\infty}(-i\kappa g)^q\sum_{(\mathcal{A}_i)_{1\leq n\leq r+q}\in\mathcal{O}_{c,q}}\underset{\sum_i\tilde{\tau}_i\leq t}{\iint\left(\prod_{i=1}^{c+1+q}d\tilde{\tau}_{i}\right)}\mathcal{P}_{\rm{E}}\mathcal{G}\left(t-{\tiny \sum}_i\tilde{\tau}_i\right)\mathcal{L}_{\rm{n}}\mathcal{G}_0(\tilde{\tau}_{c+q+1})\prod_{i=1}^{c+q}\left(\mathcal{A}_{i}\mathcal{G}_0(\tilde{\tau}_{i})\right)\rho(0),
				\end{multline}
				where we defined the set
				\begin{equation}
				\mathcal{O}_{k,q}=\left\{(\mathcal{A}_j)_{1\leq j\leq k+q}| \text{ with }\mathcal{A}_{j}\in\{ \mathcal{H} ,\, \mathcal{L}_{\rm{n}}\}\text{ and } \mathcal{A}_{j}=\mathcal{L}_{\rm{n}} \text{ for exactly } k \text{ values of } j\in\{1,..., k+q\}\right\}.
				\end{equation}
				Moreover, one remarks that
				\begin{multline}
					\sum_{q=0}^{+\infty}(-i\kappa g)^q\sum_{(\mathcal{A}_i)\in\mathcal{O}_{k,q}}\underset{\sum_i \tau_i\leq t}{\iint\left(\prod_{i=1}^{k+q}d\tau_{i}\right)}\mathcal{G}_0\left(t-{\tiny \sum}_i\tau_i\right)\prod_{i=1}^{k+q}\left(\mathcal{A}_{i}\mathcal{G}_0(\tau_i)\right)=\\
					\sum_{q=0}^{+\infty}(-i\kappa g)^q\sum_{(\mathcal{A}_i)\in\mathcal{O}_{k-1,q}}\underset{\sum_i \tau_i\leq t}{\iint\left(\prod_{i=1}^{k+q}d\tau_{i}\right)}\mathcal{G}_0\left(t-{\tiny \sum}_i\tau_i\right)\prod_{i=1}^{k+q-1}\left(\mathcal{A}_{i}\mathcal{G}_0(\tau_{i+1})\right)\mathcal{L}_{\rm{n}}\mathcal{G}_1(\tau_1)\label{SM_eq:resummation_first_H}
				\end{multline}
				This is obtained by resummating  all the Hamiltonian terms in the left handsite of the previous identity (starting from the first terms on the rightside of the product) for each $(\mathcal{A}_i)$ until the first index $i_\mathcal{A}$ such that $\mathcal{A}_{i_\mathcal{A}}=\mathcal{L}_{\rm{n}}$. This leads to the following expansion for the full projected propagator:
				\begin{multline}
					\mathcal{P}_{\rm{E}}\rho(t)-e^{-i\mathcal{H}_0 t}\rho(0)=\sum_{k=1}^{c}\kappa^k\sum_{q=0}^{+\infty}(-i\kappa g)^q\sum_{(\mathcal{A}_i)\in\mathcal{O}_{k-1,q}}\underset{\sum_i \tau_i\leq t}{\iint\left(\prod_{i=1}^{k+q}d\tau_{i}\right)}\mathcal{P}_{\rm{E}}\mathcal{G}_0\left(t-{\tiny \sum}_i\tau_i\right)\prod_{i=1}^{k+q}\left(\mathcal{A}_{i}\mathcal{G}_0(\tau_{i+1})\right)\mathcal{L}_{\rm{n}}\mathcal{G}_1(\tau_1)\rho(0)\\
					+\kappa^{c+1}\sum_{q=0}^{+\infty}(-i\kappa g)^q\sum_{(\mathcal{A}_i)\in\mathcal{O}_{c-1,q}}\underset{\sum_i \tau_i\leq t}{\iint\left(\prod_{i=1}^{k+q}d\tau_{i}\right)}\mathcal{P}_{\rm{E}}\mathcal{G}\left(t-{\tiny \sum}_i\tilde{\tau}_i\right)\mathcal{L}_{\rm{n}}\mathcal{G}_0(\tilde{\tau}_{c+q+1})\prod_{i=1}^{k+q}\left(\mathcal{A}_{i}\mathcal{G}_0(\tau_{i+1})\right)\mathcal{L}_{\rm{n}}\mathcal{G}_1(\tau_1)\rho(0),
				\end{multline}
				The first line in the right handside of  this expansion is actually vanishing. Indeed, by identification between the expansion of $\mathcal{G}\left(t\right)$ in powers of $\mathcal{L}_{\rm{n}}$ and $\mathcal{H}$ in both the Schrodinger and interaction pictures we can show that
				\begin{eqnarray}
				\sum_{k_1}\frac{(R\kappa )^{k_1}t^{k_1+q+k}}{(k_1+q+k)!}\mathcal{S}[ \mathcal{L}_{\rm{E}},\mathcal{H},\mathcal{L}_{\rm{n}},k_1,q,k]=
				\sum_{(\mathcal{A}_i)_{1\leq n\leq k+q}\in\mathcal{O}_{k,q}}\underset{\sum_i \tau_i\leq t}{\iint\left(\prod_{i=1}^{k+q}d\tau_{i}\right)}\mathcal{G}_0\left(t-{\tiny \sum}_i\tau_i\right)\prod_{i=1}^{k+q}\left(\mathcal{A}_{i}\mathcal{G}_0(\tau_i)\right).
				\end{eqnarray}
				for all integers $k,q$. Combining this identification with Lemma hypothesis (2) and the identity Eq.~(\ref{SM_eq:resummation_first_H}) one gets
				\begin{equation}
				\sum_{q=0}^{+\infty}(-i\kappa g)^q\sum_{(\mathcal{A}_i)\in\mathcal{O}_{k-1,q}}\underset{\sum_i \tau_i\leq t}{\iint\left(\prod_{i=1}^{k+q}d\tau_{i}\right)}\mathcal{P}_{\rm{E}}\mathcal{G}_0\left(t-{\tiny \sum}_i\tau_i\right)\prod_{i=1}^{k+q}\left(\mathcal{A}_{i}\mathcal{G}_0(\tau_{i+1})\right)\mathcal{L}_{\rm{n}}\mathcal{G}_1(\tau_1)\rho(0)=0.\label{SM_eq:relation_AQEC_interaction}
				\end{equation}
				for $k\in\{1,...,c\}$. The latter identity can exploited to show furthermore that:
				
				\begin{multline}
					\kappa^{c+1}\sum_{q=0}^{+\infty}(-i\kappa g)^q\sum_{(\mathcal{A}_i)\in\mathcal{O}_{c-1,q}}\underset{\sum_i \tau_i\leq t}{\iint\left(\prod_{i=1}^{k+q}d\tau_{i}\right)}\mathcal{P}_{\rm{E}}\mathcal{G}\left(t-{\tiny \sum}_i\tau_i\right)\mathcal{L}_{\rm{n}}\mathcal{G}_0(\tau_{c+q+1})\prod_{i=1}^{k+q}\left(\mathcal{A}_{i}\mathcal{G}_0(\tau_{i+1})\right)\mathcal{L}_{\rm{n}}\mathcal{G}_1(\tau_1)\rho(0)=\\
					\kappa^{c+1}\sum_{q=0}^{+\infty}(-i\kappa g)^q\sum_{(\mathcal{A}_i)\in\tilde{\mathcal{O}}_{c-1,q}}\underset{\sum_i \tau_i\leq t}{\iint\left(\prod_{i=1}^{k+q}d\tau_{i}\right)}\mathcal{P}_{\rm{E}}\mathcal{G}\left(t-{\tiny \sum}_i\tau_i\right)\mathcal{L}_{\rm{n}}\mathcal{G}_0(\tau_{c+q+1})\prod_{i=1}^{k+q}\left(\mathcal{A}_{i}\mathcal{G}_0(\tau_{i+1})\right)\mathcal{L}_{\rm{n}}\mathcal{G}_1(\tau_1)\rho(0).\label{SM_eq:identity_Q_insert}
				\end{multline}
				where we defined in the right handside the modified set
				\begin{equation}
				\tilde{\mathcal{O}}_{k,q}=\left\{(\mathcal{A}_j)_{1\leq j\leq k+q}|\text{ with } \mathcal{A}_{j}\in\{ \mathcal{Q}_{\rm{E}}\mathcal{H}\mathcal{Q}_{\rm{E}} , \mathcal{Q}_{\rm{E}}\mathcal{L}_{\rm{n}}\mathcal{Q}_{\rm{E}}\}\text{ and } \mathcal{A}_{j}=\mathcal{Q}_{\rm{E}}\mathcal{L}_{\rm{n}}\mathcal{Q}_{\rm{E}} \text{ for exactly } k \text{ values of } j\in\{1,..., k+q\}\right\}.
				\end{equation}
				The relation of Eq.~(\ref{SM_eq:identity_Q_insert})  can be obtained recursively using Eq.~(\ref{SM_eq:relation_AQEC_interaction})  by inserting the identity $\mathcal{I}=\mathcal{P}_{\rm{E}}+\mathcal{Q}_{\rm{E}}$ between each $\mathcal{L}_{\rm{n}}$ or $\mathcal{A}_i$ and each $\mathcal{G}_0(\tau_{i})$ in the left handside of Eq.~(\ref{SM_eq:identity_Q_insert}).
				This leads to
				\begin{multline}
					\mathcal{P}_{\rm{E}}\rho(t)-e^{-i\mathcal{H}_0 t}\rho(0)=\kappa^{c+1}\sum_{q=0}^{+\infty}(-i\kappa g)^q\sum_{(\mathcal{A}_i)\in\tilde{\mathcal{O}}_{c-1,q}}\underset{\sum_i \tau_i\leq t}{\iint\left(\prod_{i=1}^{k+q}d\tau_{i}\right)}\\\mathcal{P}_{\rm{E}}\mathcal{G}\left(t-{\tiny \sum}_i\tau_i\right)\mathcal{L}_{\rm{n}}\mathcal{G}_0(\tau_{c+q+1})\prod_{i=1}^{k+q}\left(\mathcal{A}_{i}\mathcal{G}_0(\tau_{i+1})\right)\mathcal{L}_{\rm{n}}\mathcal{G}_1(\tau_1)\rho(0).
				\end{multline}
				
				By resummation of the previous expansion in powers of the projected Hamiltonian superoperator $\mathcal{Q}_{\rm{E}}\mathcal{H}\mathcal{Q}_{\rm{E}}$ the last identity can be rewritten as:
				\begin{equation}
				\mathcal{P}_{\rm{E}}\rho(t)-e^{-i\mathcal{H}_0 t}\rho(0)=\kappa^{c+1}\underset{\sum_i \tau_i\leq t}{\iint\left(\prod_{i=1}^{c+1}d\tau_{i}\right)}\mathcal{P}_{\rm{E}}\mathcal{G}\left(t-{\tiny \sum}_i\tau_i\right)\mathcal{L}_{\rm{n}}\mathcal{G}_1^{\mathcal{Q}_{\rm{E}}}(\tau_{c+1})\mathcal{Q}_{\rm{E}}\prod_{i=2}^{c}\left(\mathcal{L}_{\rm{n}}\mathcal{G}_1^{\mathcal{Q}_{\rm{E}}}(\tau_i)\mathcal{Q}_{\rm{E}}\right)\mathcal{L}_{\rm{n}}\mathcal{G}_1(\tau_1)\rho(0).\label{SM_eq:resummation_proj_ham}
				\end{equation}
				with $\mathcal{G}_1^{\mathcal{Q}_{\rm{E}}}(\tau)=\text{exp}[(\kappa R\mathcal{L}_{\rm{E}}-i\kappa g  \mathcal{Q}_{\rm{E}}\mathcal{H}\mathcal{Q}_{\rm{E}})t]$ (we used the fact that $\mathcal{L}_{\rm{E}}=\mathcal{Q}_{\rm{E}}\mathcal{L}_{\rm{E}}=\mathcal{L}_{\rm{E}}\mathcal{Q}_{\rm{E}}$). To prove the desired bound, we will use the following property:
				\begin{center}
					\emph{Exist $g_0,N,\lambda_0>0$, such that for all $R>0$ and for all $\kappa g\in\mathbb{R}$ such that $|\kappa g|<g_0 R\kappa$  one has}
					\begin{equation}
					\left\|\mathcal{G}_1^{\mathcal{Q}_{\rm{E}}}(t)\mathcal{Q}_{\rm{E}}\right\|\leq M e^{-R\kappa \lambda_0 t}\label{SM_eq:relaxation_Hamiltonian}
					\end{equation}
				\end{center}
				The bound Eq.~(\ref{SM_eq:relaxation_Hamiltonian}) can be proved by expanding this $\mathcal{G}_1^{\mathcal{Q}_{\rm{E}}}(t)\mathcal{Q}_{\rm{E}}$ in powers of  $ \mathcal{Q}_{\rm{E}}\mathcal{H}\mathcal{Q}_{\rm{E}}$:
				\begin{equation}
				\mathcal{G}_1^{\mathcal{Q}_{\rm{E}}}(t)\mathcal{Q}_{\rm{E}}=\mathcal{G}_0(t)\mathcal{Q}_{\rm{E}}+\sum_{k=1}^{+\infty}\underset{\sum_i \tau_i\leq t}{\iint\left(\prod_{i=1}^{k}d\tau_{i}\right)}\mathcal{G}_0(t-\sum \tau_i)\mathcal{Q}_{\rm{E}}\prod_{i=1}^{k}\left(\kappa g \mathcal{Q}_{\rm{E}}\mathcal{H}\mathcal{Q}_{\rm{E}}\mathcal{G}_0(\tau_i)\right)\mathcal{Q}_{\rm{E}}.
				\end{equation}
				Since $\mathcal{Q}_{\rm{E}}$ projects on relaxation eigenmodes of $\mathcal{L}_{\rm{E}}$, we deduce that there exists $\tilde{M}>0$ such that $\mathcal{G}_0(\tau_i)\mathcal{Q}_{\rm{E}}\leq \tilde{M}e^{-R\alpha\kappa\tilde{\lambda}_0 t}$ for all $R,\kappa,t>0$, $0<\alpha\leq 1$ and $\tilde{\lambda}_0=-\text{Max}[\text{Re}(\textrm{Sp}(\mathcal{L}_{\rm{E}}))\backslash\{0\}]>0$ is the real part of the eigenvalue corresponding to the slowest relaxation eigenmode of $\mathcal{L}_{\rm{E}}$. From this we get:
				\begin{eqnarray}
				\left\|\mathcal{G}_1^{\mathcal{Q}_{\rm{E}}}(t)\mathcal{Q}_{\rm{E}}\right\|&\leq& \tilde{M}e^{-R\kappa\alpha\tilde{\lambda}_0t}+\sum_{k=1}^{+\infty}\underset{\sum_i \tau_i\leq t}{\iint\left(\prod_{i=1}^{k}d\tau_{i}\right)}\tilde{M}^{k+1} e^{-R\kappa\alpha\tilde{\lambda}_0t}(|\kappa g|\|\mathcal{Q}_{\rm{E}}\mathcal{H}\mathcal{Q}_{\rm{E}}\|)^k\\
				&=&\tilde{M}e^{-R\kappa\alpha\tilde{\lambda}_0t}\sum_{k=0}^{+\infty} \frac{(\tilde{M}|\kappa g|\|\mathcal{Q}_{\rm{E}}\mathcal{H}\mathcal{Q}_{\rm{E}}\|t)^k}{k!}\\
				&=&\tilde{M}e^{-(R\kappa\alpha\tilde{\lambda}_0-\tilde{M}|\kappa g|\|\mathcal{Q}_{\rm{E}}\mathcal{H}\mathcal{Q}_{\rm{E}}\|)t}
				\end{eqnarray}
				where we used the identity 
				$$\underset{\sum_i \tau_i\leq t}{\iint\left(\prod_{i=1}^{k}d\tau_{i}\right)}1=t^k/k!$$ 
				for the integral over regular triangular domain of dimension $k$. We see that for  $|\kappa g|\leq g_0 R\kappa$ with $g_0=\alpha\tilde{\lambda}_0/(2\tilde{M}\|\mathcal{Q}_{\rm{E}}\mathcal{H}\mathcal{Q}_{\rm{E}}\|)$ we get the expected bound with $M=\tilde{M}$ and $\lambda_0=\alpha\tilde{\lambda}_0/2$. 
				
				Inserting the bound Eq.~(\ref{SM_eq:relaxation_Hamiltonian}) in the identity Eq.~(\ref{SM_eq:resummation_proj_ham}) and using the fact that the first propagator $\mathcal{G}_1(\tau_1)$ in the expansion of Eq.~(\ref{SM_eq:resummation_proj_ham}) is a CPTP map and thus is bounded by a time-independent constant we find that there exists $M>0$ such that the following bound is verified for all $\rho(0)\in\mathbb{C}$, for all $\kappa,R,t\geq  0$ and $g\in\mathbb{R}$ such that $|\kappa g|<g_0 R\kappa$:
				\begin{equation}
				\left\|\mathcal{P}_{\rm{E}}\rho(t)-e^{-i\kappa g\mathcal{H}_0 t}\rho(0)\right\|\leq \frac{N\kappa t}{R^c}\|\rho(0)\|
				\end{equation}
				which is the desired result.
				\section{Some generalized error-transparent Hamiltonians}
				In Eq.~(\jose{8}) of the main manuscript we see that our specific explicit construction for the Hamiltonian  $H$ satisfies both conditions $HEP_{\mathcal{C}}=EH_0P_{\mathcal{C}}$ and $[H,E]P_{\mathcal{C}}=0$, for all the error operators  $E\in\mathcal{E}^{[\sim c]}$  of our error set, coinciding thus exactly with the most standard definitions of the  error-transparent Hamiltonian \cite{ETH_Vy,ETH}. However it is possible to show that  the condition (2) of  the  Lemma admits more solutions and that not all possible generalized Hamiltonians can be cast in this form. We outline here two constructions (without going through the derivation) going beyond the framework of the standard error-transparent Hamiltonian.
				\begin{itemize} 
					\item \textbf{A generalized ETH which does not not preserve  the error syndrome or commute with the errors.} Standard error-transparent gates have the property that a natural error happening before the gate operation produces the same final state as the error happening immediately after the gate.  Intuitively one expects however a gate generated by an Hamiltonian $H$ producing various final states (depending on when the natural dissipation error happened) with the \emph{same} logical content but differing errors syndromes to be still functional as long as the various generated errors have the same weight: under those conditions the robustness of the code is not hindered by the Hamiltonian and an autonomously error-corrected computation of order $c$ should still be achievable. 
					
					In that perspective, a simple example of viable modification to $H$ would be to add to the Hamiltonian in Eq.~(\jose{8}) of the main  manuscript an extra contribution of the form $\Delta H=\sum_{k=1}^{d_{\mathcal{C}}}\sum_{n=0}^{c}\sum_{i_{n},j_{n}=1}^{p_{n}}E_{i_{n},j_{n}}^{n}|\mu_{k,i_{n}}^{[n]}\rangle\langle\mu_{k,j_{n}}^{[n]}|+h.c.$ coupling states corresponding to different  error syndromes, but with identical error weight and underlying logical word, via  a coupling strength $E_{i_{n},j_{n}}^{n}$ independent of  the code word. With respect to our explicit construction $\mathcal{L}_{\rm{E}}$ does need to be modified. 
					Since $\Delta H$ does not discriminate between the various code states and their associated correctable error-states ($E_{i_{n},j_{n}}^{n}$ does not depend on the code word), such correction to $H$ does not impact at all the logical computation. Moreover since  $H$ does not affect  the error weight, it does not fragilize further the code against natural dissipation error, and the  protection up to order $c$ is preserved.  
					
					The fact that the modified Hamilotnian $H$ does not preserves the error syndrome given by the indices  $n$, $i_{n}$  of the states $|\mu_{k,i_{n}}^{[n]}\rangle$ is connected to a non-commutativity with the error model: to understand this, let us consider  a toy model composed of a 2-dimensional code embedded in a 6-dimensional Hilbert space. A basis of this Hilbert space is given by two code space states $\{\ket{\mu},\ket{\nu}\}$, and 4 error states $\{\ket{\mu_1},\ket{\mu_2},\ket{\nu_1},\ket{\nu_2}\}$. We consider the natural dissipation model corresponding to the set of jumps $\{F_a,\,a=1,2\}$  with $F_a=\ket{\mu_a}\bra{\mu}+\ket{\nu_a}\bra{\nu}+\ket{\mu_a}\bra{\nu_a}+\ket{\nu_a}\bra{\mu_a}$. It is simple to show that the set 
					$\mathcal{E}^{[\sim 1]}=\{\mathbb{1},F_1,F_2\}$ satisfies the Knill-Laflame conditions and thus this model is suitable for 1st-order AutoQEC. Moreover, as it is introduced, the above Hilbert space basis is already a naturally Gram-Schmidt orthonomalized error state basis which can  be used for the construction of our engineered dissipation and generalized ETH in Eqs.~(\jose{7-8}) of the main  manuscript. Let us now consider the following error-syndrome modifying additive contribution
					$\Delta H=\ket{\mu_1}\bra{\mu_2}+\ket{\nu_1}\bra{\nu_2}+h.c$ to the generalized ETH $H$  in Eq.~(\jose{8}) of the main  manuscript. One finds that it does not satisfy the comutation relation of a standard ETH with the error set $\Delta H F_1 P_\mathcal{C}=F_2\Delta HP_\mathcal{C}\neq F_1\Delta HP_\mathcal{C} $ and  $\Delta H F_2 P_\mathcal{C}=F_1\Delta HP_\mathcal{C}\neq F_2\Delta HP_\mathcal{C}$. This is directly related to an exchange of the error syndromas $a=1,2$ upon application of  $\Delta H$.
					
					\item \textbf{A generalized ETH which does not preserve  the code space:} Interestingly, $H$ does not even have to preserve the code space.  Indeed, let us hypothetically extend the Hilbert space by adding a copy $\tilde{\mathcal{H}}$ of $\mathcal{H}$.
					The copy space $\tilde{\mathcal{H}}$ consists of a copy $\tilde{\mathcal{C}}=\textrm{span}\{\ket{\tilde{\mu}_i},1\leq i \leq d_{\mathcal{C}}\}$ of the code space $\mathcal{C}$, in  direct sum with a copy of the space of correctable states ($\textrm{span}\{\ket{\tilde{\mu}}_{k,i_{n}}^{[n]},i_n=1..p_n|1\leq n\leq c,1\leq k\leq d_{\mathcal{C}}\}$) and a copy of the space of residual states  ($\textrm{span}\{\lbrace |\tilde{\phi}_{q}\rangle,1\leq q\leq q_{\textrm{max}}\}$). We assume that $\tilde{\mathcal{H}}$ is free of any natural dissipation and that $\mathcal{H}$ and $\tilde{\mathcal{H}}$ do not couple via natural dissipation. It is then possible to show that the following engineered dissipation and generalized ETH are suitable for error-corrected quantum computations of order $c$:
					\begin{eqnarray}
					\mathcal{L}_{\textrm{E}} &=& \sum_{n=1}^{c}\sum_{i_{n}=1}^{p_{n}} D[ F_{\textrm{E},i_{n}}^{[n]} ] + \sum_{q=1}^{q_{\max}} D[F_{\textrm{E},q}^{[\textrm{res}]}]+\sum_{n=1}^{c}\sum_{i_{n}=1}^{p_{n}} D[ \tilde{F}_{\textrm{E},i_{n}}^{[n]} ] + \sum_{q=1}^{q_{\max}} D[\tilde{F}_{\textrm{E},q}^{[\textrm{res}]}]\\
					H&=&\sum_{j,k=1}^{d_{\mathcal{C}}}\sum_{n=0}^{c}\sum_{i_{n}=1}^{p_{n}} \langle\mu_j|H_0|\mu_k\rangle|\mu_{j,i_{n}}^{[n]}\rangle\langle\mu_{k,i_{n}}^{[n]}|+
					\sum_{j,k=1}^{d_{\mathcal{C}}}\sum_{n=0}^{c}\sum_{i_{n}=1}^{p_{n}} \langle\mu_j|H_0|\mu_k\rangle|\tilde{\mu}_{j,i_{n}}^{[n]}\rangle\langle\tilde{\mu}_{k,i_{n}}^{[n]}|\nonumber\\
					&&+\sum_{k=0}^{d_{\mathcal{C}}} \left[|\tilde{\mu}_k\rangle\langle\mu_k|+h.c\right]+\sum_{k=0}^{d_{\mathcal{C}}}\sum_{n=1}^{c}\sum_{i_{n}=1}^{p_{n}} \left[|\tilde{\mu}_{k,i_{n}}^{[n]}\rangle\langle\mu_{k,i_{n}}^{[n]}|+h.c\right]+\sum_{q=1}^{q_{\textrm{max}}} \left[ |\tilde{\phi}_{q}\rangle\langle\phi_{q}|+h.c\right].\label{eq:mod_H}
					\end{eqnarray}
					From the second line in Eq.~(\ref{eq:mod_H}) we see that those contributions to $H$ completely swap the Hilbert spaces $\mathcal{H}$ and $\tilde{\mathcal{H}}$: in particular the code space $\mathcal{C}$ is not stable under application of $H$ as it couples to $\tilde{\mathcal{C}}$ and one finds $HP_{\mathcal{C}}\neq H_0P_{\mathcal{C}}$. The second contribution in the first line  Eq.~(\ref{eq:mod_H}) ensures that the computation is still `error-transparent' in the copy space $\tilde{\mathcal{H}}$. To ensure that the probabibilty converges back to the orginal Hilbert space and eventually the code space, the engineered dissipation is completed with respect to our original construction by an extra series of jumps of the form $ \tilde{F}_{\textrm{E},i_{n}}^{[n]}= \sum_{j=1}^{d_{\mathcal{C}}}| \mu_{j,i_{n}}^{[n]}\rangle\langle \tilde{\mu}_{j,i_{n}}^{[n]}|$ and $\tilde{F}_{\textrm{E},q}^{[\textrm{res}]} =| \phi_{q}\rangle \langle \tilde{\phi}_{q}|$. 
					
					Physically speaking, a copy of such Hilbert space and the associated swapping induced by the Hamiltonian $H$ is straightforwardly obtained by adding  a two-level and an extra ancilla, which Rabi oscillates between its fundamental and excited states via a $\sigma_x$ contribution in $H$. Such generalized ETH associated with the additional ancilla is related to subsystem code, with the extra ancilla plays the role of gauge degrees of freedom \cite{Bacon_subsystem_QEC}. While this would require further verification, our intuition is that a single extra engineered jump operator of the form $\tilde{F}_{\textrm{E}}= \sigma_-$ for the extra ancilla might as well work instead of having an extra jump ($ \tilde{F}_{\textrm{E},i_{n}}^{[n]}$ and $\tilde{F}_{\textrm{E},q}^{[\textrm{res}]} $) for each error syndrome and residual state.
				\end{itemize}
				
				\section{Effective logical decoherence dynamics}
				Let us have some engineered dissipation $\mathcal{L}_{\rm{E}}$ and an Hamiltonian $H$ performing an error-corrected autonomous quantum computation up  to order $c$ wrt the code space $\mathcal{C}$  and the natural dissipation $\mathcal{L}_{\textrm{n}}$. In particular, due to the Lemma, the recovery projector satisfies $\mathcal{P}_{\rm{E}}\mathcal{P}_{\mathcal{C}}=\mathcal{P}_{\mathcal{C}}$. As stated in the main manuscript we also assume $\mathcal{P}_{\mathcal{C}}\mathcal{P}_{\rm{E}}=\mathcal{P}_{\rm{E}}$.
				\subsection{Projected density matrix}
				$\mathcal{P}_{\rm{E}}$ being a projector one can apply Nakajima-Zwanzig projection operator techniques \cite{Breuer_projective}:  assuming an  initial condition $\rho(0)$ in the code space (i.e. satisfying $\mathcal{P}_{\rm{E}}\rho(0)$ since $\mathcal{P}_{\mathcal{C}}\mathcal{P}_{\rm{E}}=\mathcal{P}_{\rm{E}}$), the recovered density matrix $\mathcal{P}_{\rm{E}}\rho(t)$ follows the exact non-Markovian dynamics
				\begin{equation}
				\partial_t\mathcal{P}_{\rm{E}}\rho=\mathcal{P}_{\rm{E}}\mathcal{L}\mathcal{P}_{\rm{E}}+\int_0^t  d\tau \Sigma(\tau)\mathcal{P}_{\rm{E}}\rho(t-\tau).\label{eq:Nakajima-Zwanzig}
				\end{equation}
				where the memory kernel is defined as 
				\begin{equation}
				\Sigma(\tau)\equiv\mathcal{P}_{\rm{E}}\mathcal{L}\mathcal{Q}_{\rm{E}}\textrm{exp}[\mathcal{Q}_{\rm{E}}\mathcal{L}\mathcal{Q}_{\rm{E}}\tau]\mathcal{Q}_{\rm{E}}\mathcal{L}\mathcal{P}_{\rm{E}}
				\end{equation}
				and represents the sum of all processes  leaving the projective space associated to $\mathcal{P}_{\rm{E}}$ (in our case the code space) evolving for a time $\tau$ in the complementary space associated to the projector $\mathcal{Q}_{\rm{E}}$ (here the relaxation modes of engineered dissipation) and coming back finally in the projective space. Importantly as a consequence of our  hypothesis $\mathcal{P}_{\mathcal{C}}\mathcal{P}_{\rm{E}}=\mathcal{P}_{\rm{E}}$, the memory kernel can be rewritten as
				\begin{equation}
				\Sigma(\tau)\equiv\mathcal{P}_{\rm{E}}\mathcal{L}\mathcal{Q}_{\rm{E}}\textrm{exp}[\mathcal{Q}_{\rm{E}}\mathcal{L}\mathcal{Q}_{\rm{E}}\tau]\mathcal{Q}_{\rm{E}}\mathcal{L}\mathcal{P}_{\mathcal{C}}\mathcal{P}_{\rm{E}}.
				\end{equation}
				The insertion of a code space projector $\mathcal{P}_{\mathcal{C}}$ on the right hand side plays a key role in our  derivation, as it can then be shown that  expanding $\Sigma(\tau)$ in function of the various processes $\mathcal{L}_{\rm{n}}$, $\mathcal{L}_{\rm{E}}$ and $\mathcal{H}$ will lead to the cancellation of all $c+1$ lowest-order contributions in powers of $\mathcal{L}_{\rm{n}}$ via the property (2) of the Lemma. Remarking that $\mathcal{P}_{\rm{E}}\mathcal{L}\mathcal{P}_{\rm{E}}=\mathcal{P}_{\rm{E}}\mathcal{L}\mathcal{P}_{\mathcal{C}}\mathcal{P}_{\rm{E}}=-ig\kappa\mathcal{H}_0\mathcal{P}_{\rm{E}}$,
				the master equation Eq.~(\ref{eq:Nakajima-Zwanzig}) can be rewritten as:
				\begin{equation}
				\partial_t\mathcal{P}_{\rm{E}}\rho=-ig\kappa\mathcal{H}_0+\mathcal{L}_{\rm{eff}}\mathcal{P}_{\rm{E}}\rho(t)+\left[\partial_t\mathcal{P}_{\rm{E}}\rho\right]_{\rm{nloc}}+\left[\partial_t\mathcal{P}_{\rm{E}}\rho\right]_{\rm{kink}}
				\end{equation}
				where the effective Liouvillian is given by
				\begin{equation}
				\mathcal{L}_{\rm{eff}}=\int_{0}^{+\infty} d\tau\left[\Sigma(\tau)e^{i g\kappa\mathcal{H}_0\tau}\right]\mathcal{P}_{\mathcal{C}}\label{eq:effective_liouvillian}
				\end{equation}
				The corrections
				\begin{eqnarray}
				\left[\partial_t\mathcal{P}_{\rm{E}}\rho\right]_{\rm{nloc}}&=&\int_0^td\tau \Sigma(\tau)\left[\mathcal{P}_{\rm{E}}\rho(t-\tau)-e^{i g\kappa\mathcal{H}_0\tau}\mathcal{P}_{\rm{E}}\rho(t)\right]\label{SM_eq:nloc_contrib}\\
				\left[\partial_t\mathcal{P}_{\rm{E}}\rho\right]_{\rm{kink}}&=&\int_t^{+\infty}d\tau \Sigma(\tau)e^{i g\kappa\mathcal{H}_0\tau}\mathcal{P}_{\rm{E}}\rho(t).\label{SM_eq:kink_contrib}
				\end{eqnarray}
				correspond to memory-related non-local effects and the initial kink of the dynamics.  
				
				Let us now have $\tilde{g_0}>0$ so that Eq.~(\ref{SM_eq:relaxation_Hamiltonian})  in the supplementary is satisfied for all $R>0$ and $|g|\leq \tilde{g_0}R$. On one hand,  the memory kernel $\Sigma(\tau)$ is expected to decay exponentially to zero over the short time scale $\sim1/(R\kappa)$ for a large enough engineered dissipation strength $R$ and a controlled Hamiltonian  coupling $|g|\leq \tilde{g_0} R$. This is because $\Sigma(\tau)$ contains as only time-dependent factor the term $\textrm{exp}[\mathcal{Q}_{\rm{E}}\mathcal{L}\mathcal{Q}_{\rm{E}}\tau]\mathcal{Q}_{\rm{E}}$: after expanding such term in powers of $\mathcal{L}_{\rm{n}}$, it is easy to prove this result by using the fact that $\mathcal{G}_1^{\mathcal{Q}_{\rm{E}}}(\tau)\mathcal{Q}_{\rm{E}}=\text{exp}[(\kappa R\mathcal{L}_{\rm{E}}-i\kappa g  \mathcal{Q}_{\rm{E}}\mathcal{H}\mathcal{Q}_{\rm{E}})t]\mathcal{Q}_{\rm{E}}$ also decays over  over the time scale $\sim1/(R\kappa)$ as  a consequence of  Eq.~(\ref{SM_eq:relaxation_Hamiltonian}). 
				
				On the other hand, since $H$ and  $\mathcal{L}_{\rm{E}}$ perform autonomous error-corrected quantum computation  up  to order $c$, one expects that beyond a simple time-dependent and deterministic unitary rotation $U_0(t)=\textrm{exp}[-ig\kappa H_0t]$ that the relaxation dynamics of $\mathcal{P}_E\rho(t)$ should be very slow (occuring over the time scale $t\sim  R^c/\kappa$ as soon as $R\gg1$), and thus the  memory kernel  should be suppressed for large $R$. More precisely, using the additional hypothesis $\mathcal{P}_{\mathcal{C}}\mathcal{P}_{\rm{E}}=\mathcal{P}_{\rm{E}}$ and the fact that $\mathcal{L}_{\rm{E}}$ and $H$ satisfies the lemma condition (2), one  proves that all $c+1$ leading orders in the expansion of the memory kernel  in powers of natural dissipation cancel out exactly. 
				
				Ultimately,  similarily to was done in Sec.~\ref{sec:2_1_lemma} of  the supplementary, the kernel  can be bounded as $\|\Sigma(\tau)\|\leq A\kappa^2\textrm{exp}[-R\lambda_0\kappa  \tau]/R^{c-1}$ for all $\tau,R\geq0$ and all $g\leq \tilde{g_0}R$  for some dimensionless constants $A,\lambda_0>0$. Combined with Eq.~(\ref{eq:Nakajima-Zwanzig}), this bound implies that the relaxation dynamics of $\mathcal{P}_{\rm{E}}\rho(t)$ must be slow, which leads the following bound
				\begin{equation}
				\left\|\mathcal{P}_E\rho(t)-U_0(g\kappa(t-t'))[\mathcal{P}_E\rho(t')]U_0^{\dagger}(g\kappa(t-t'))\right\|
				\leq \tilde{M}\kappa|t-t'|/R^c
				\end{equation}
				for some $\tilde{M}>0$, for all $R,t,t'\geq0$ and $g$ such that $|g|\leq   \tilde{g}_0R $, which is a generalization of Eqs.~(\jose{3-4}) of the  main manuscript.
				
				As a consequence of this  very slow decoherence and the fast relaxation of $\Sigma(\tau)$, one expects thus that non time-local features of the Eq.~(\ref{eq:Nakajima-Zwanzig}) quantified by the correction Eq.~(\ref{SM_eq:nloc_contrib}) should be negligible. Similarly, the kink term in Eq.~(\ref{SM_eq:kink_contrib}) should only have some limited impact restrained to the very early stage of the dynamics due to the fast relaxation of the memory kernel. A precise analysis yields the exact upper bounds:
				\begin{eqnarray}
				\left[\partial_t\mathcal{P}_{\rm{E}}\rho\right]_{\rm{nloc}}&\leq& \tilde{M}\frac{\kappa}{R^{2c+1}}\|\rho(0)\|\\
				\left[\partial_t\mathcal{P}_{\rm{E}}\rho\right]_{\rm{kink}}&\leq& \tilde{N}\frac{\kappa}{R^c}e^{-\beta\kappa Rt}\|\rho(0)\|
				\end{eqnarray}
				for some positive dimensionless numbers $\tilde{M},\tilde{N},\beta>0$ for all $\rho(0)\in\mathcal{C}\otimes_{\rm{d}}\mathcal{C}$ and $R,\kappa,t\geq 0$, $g$ satisfying $|g|\leq \tilde{g}_0R$. Considering that $\partial_t\mathcal{P}_{\rm{E}}\rho\sim \kappa/R^c$, one finds thus that the the non-local contribution $\left[\partial_t\mathcal{P}_{\rm{E}}\rho\right]_{\rm{nloc}}$ can be completely neglected in the limit $R\to+\infty$. The kink contribution $\left[\partial_t\mathcal{P}_{\rm{E}}\rho\right]_{\rm{kink}}$ is relevant at very short times $t\simeq 1/(\kappa R)$, but since it decays immediately afterwards, it leads to overall negligibe deviations $\mathcal{O}\left(1/R^{c+1}\right)$ in the evolution of the density matrix.

				\subsection{Full density matrix}
				Our formalism enables us to retrieve information about the full density matrix $\rho(t)$ from our knowledge of the dynamics of the recovered matrix $\mathcal{P}_{\rm{E}}\rho(t)$ : if $\rho(0)\in\mathcal{C}\otimes\mathcal{C}$ then at later times ones has exactly:
				\begin{equation}
				\mathcal{Q}_{\rm{E}}\rho(t)=\int_0^{t}d\tau e^{\mathcal{Q}_{\rm{E}}\mathcal{L}\mathcal{Q}_{\rm{E}}\tau} \mathcal{P}_{\rm{E}}\rho(t-\tau).
				\end{equation}
				Proceeding similarly to the previous section one can show that $\mathcal{Q}_{\rm{E}}\rho(t)=\left[\mathcal{Q}_{\rm{E}}\rho(t)\right]_{\infty}+\left[\mathcal{Q}_{\rm{E}}\rho(t)\right]_{\text{kink}}+\left[\mathcal{Q}_{\rm{E}}\rho(t)\right]_{\text{nloc}}$, where the dominant contribution  is given by
				\begin{equation}
				\left[\mathcal{Q}_{\rm{E}}\rho(t)\right]_{\infty}= \int_0^{+\infty} d\tau e^{\mathcal{Q}_{\rm{E}}\mathcal{L}\mathcal{Q}_{\rm{E}}\tau}\mathcal{Q}_{\rm{E}}\mathcal{L}e^{i g\kappa\mathcal{H}_0\tau}\mathcal{P}_{\rm{E}}\rho(t).\\
				\end{equation}
				Likewise, as a consequence of the slow logical space  decoherence dynamics associated to AutoQEC the remaining corrections are bounded as following
				\begin{eqnarray}
				\left\|\left[\mathcal{Q}_{\rm{E}}\rho(t)\right]_{\text{kink}}\right\|&\leq&\frac{A}{R}e^{-R\kappa Bt}\\
				\left\|\left[\mathcal{Q}_{\rm{E}}\rho(t)\right]_{\text{nloc}}\right\|&\leq&\frac{C}{R^{c+2}},
				\end{eqnarray}
				for all $R,t,t'\geq0$ and $g$ such that $|g|\leq  \tilde{g}_0R $, where $A,\,B,\,C$ are some positive constants. Adding together the complementary projected parts of the density matrix $\rho(t)=\mathcal{P}_{\rm{E}}\rho(t)+\mathcal{Q}_{\rm{E}}\rho(t)$, based on  the previously obtained bounds we find at all times the following estimate:
				\begin{eqnarray}
				\rho(t)&=&\mathcal{T}\mathcal{P}_{\rm{E}}\rho(t)+\delta\rho(t)
				\end{eqnarray}
				where
				\begin{equation}
				\mathcal{T}=\mathcal{I}+\int_0^{+\infty} d\tau e^{\mathcal{Q}_{\rm{E}}\mathcal{L}\mathcal{Q}_{\rm{E}}\tau}\mathcal{Q}_{\rm{E}}\mathcal{L}e^{i g\kappa\mathcal{H}_0\tau}\mathcal{P}_{\mathcal{C}}
				\end{equation}
				\begin{equation}
				\|\delta\rho(t)\|\leq \frac{A}{R}e^{-R\kappa B t}+\frac{C}{R^{c+2}}.
				\end{equation}
				\subsection{Final expressions for the effective dynamics\label{sec:final_exp}}
				Here we analyse further the expression of the superoperators $\mathcal{L}_{\rm{eff}}$ and $\mathcal{T}$ governing the dynamics of $\mathcal{P}_{\rm{E}}\rho(t)$ and $\rho(t)$. 
				\subsubsection{General case}
				Let us have a complete set $\{|\mu_v^{H_0}\rangle\in\mathcal{C}, 1\leq v\leq d_{\mathcal{C}}\}$ of code space eigenstates of the logical Hamiltonian $H_0$, and we denote $E_v$ the corresponding eigenenergies. We introduce the extra notations $|\mu_v^{H_0},\mu_w^{H_0} \rrangle=|\mu_v^{H_0}\rangle\langle\mu_w^{H_0}|$ and $E_{v,w}=E_v-E_w$.Similarily to what was done in Sec.~\ref{sec:2_1_lemma} one can expand the  memory kernel $\Sigma(\tau)$ in  powers of natural dissipation and rule out the $c-1$ lowest-order contributions since property (2) if the Lemma is satisfied. After the expansion, one can compute the required integral and one finds the following expressions: $\mathcal{L}_{\rm{eff}}=-ig\kappa\mathcal{H}_{0}+\mathcal{L}_{\rm{eff}}^{00}+\mathcal{L}_{\rm{eff}}^{10}+\mathcal{L}_{\rm{eff}}^{01}+\mathcal{L}_{\rm{eff}}^{11}$
				with 
				\begin{eqnarray}
				\llangle \mu_{t}^{H_0},\mu_u^{H_0} |\mathcal{L}_{\rm{eff}}^{00}|\mu_v^{H_0},\mu_w^{H_0} \rrangle&=&\kappa\sum_{k=c}^{+\infty}\left(\frac{-1}{R}\right)^{k}\llangle \mu_t^{H_0},\mu_u^{H_0} |\mathcal{P}_{\rm{E}}\left(\mathcal{L}_{\rm{n}}\mathcal{F}_{\rm{inv}}^{v,w}\right)^{k}\mathcal{L}_{\rm{n}}|\mu_v^{H_0},\mu_w^{H_0} \rrangle\label{eq:L_eff_exp_init}\\
				\llangle \mu_{t}^{H_0},\mu_u^{H_0} |\mathcal{L}_{\rm{eff}}^{10}|\mu_v^{H_0},\mu_w^{H_0} \rrangle&=&-ig\kappa\sum_{k=c}^{+\infty}\left(\frac{-1}{R}\right)^{k+1}\llangle \mu_t^{H_0},\mu_u^{H_0} |\mathcal{P}_{\rm{E}}\mathcal{H}\mathcal{F}_{\rm{inv}}^{v,w}\left(\mathcal{L}_{\rm{n}}\mathcal{F}_{\rm{inv}}^{v,w}\right)^{k}\mathcal{L}_{\rm{n}}|\mu_v^{H_0},\mu_w^{H_0} \rrangle\\
				\llangle \mu_{t}^{H_0},\mu_u^{H_0} |\mathcal{L}_{\rm{eff}}^{01}|\mu_v^{H_0},\mu_w^{H_0} \rrangle&=&-g\kappa\sum_{k=c}^{+\infty}\left(\frac{-1}{R}\right)^{k+1}\llangle \mu_t^{H_0},\mu_u^{H_0} |\mathcal{P}_{\rm{E}}\left(\mathcal{L}_{\rm{n}}\mathcal{F}_{\rm{inv}}^{v,w}\right)^{k+1}\mathcal{H}|\mu_v^{H_0},\mu_w^{H_0} \rrangle\\
				\llangle \mu_{t}^{H_0},\mu_u^{H_0} |\mathcal{L}_{\rm{eff}}^{11}|\mu_v^{H_0},\mu_w^{H_0} \rrangle&=&-g^2\kappa\sum_{k=c}^{+\infty}\left(\frac{-1}{R}\right)^{k+2}\llangle \mu_t^{H_0},\mu_u^{H_0} |\mathcal{P}_{\rm{E}}\mathcal{H}\mathcal{F}_{\rm{inv}}^{v,w}\left(\mathcal{L}_{\rm{n}}\mathcal{F}_{\rm{inv}}^{v,w}\right)^{k+1}\mathcal{H}|\mu_v^{H_0},\mu_w^{H_0} \rrangle\label{eq:L_eff_exp}
				\end{eqnarray}
				and
				\begin{equation}
				\mathcal{T}|\mu_v^{H_0},\mu_w^{H_0} \rrangle=\mathcal{I}+\sum_{k=1}^{+\infty}\left(\frac{-1}{R}\right)^{k}\left(\mathcal{F}_{\rm{inv}}^{v,w}\mathcal{L}_{\rm{n}}\right)^{k-1}\mathcal{F}_{\rm{inv}}^{v,w}(\mathcal{L}_{\rm{n}}-ig\mathcal{H})|\mu_v^{H_0},\mu_w^{H_0} \rrangle.
				\label{eq:T_exp}
				\end{equation}

				The superoperator $\mathcal{F}_{\rm{inv}}^{v,w}=-\int_0^{+\infty}du\textrm{exp}\left[\mathcal{L}_{\textrm{E},H}^{v,w}u\right]\mathcal{Q}_{\rm{E}}$, with
				\begin{equation}
				\mathcal{L}_{\textrm{E},H}^{v,w}=\mathcal{L}_{\rm{E}}-ig/R\left(\mathcal{Q}_{\rm{E}}\mathcal{H}\mathcal{Q}_{\rm{E}}-E_{v,w}\mathcal{Q}_{\rm{E}}\right).
				\end{equation}
				is well defined for an Hamiltonian coupling $g$ satisfying $|g|\leq \tilde{g}_0 R$, as $\tilde{g}_0$ was chosen (and proven to exist) so to satisfy the bound Eq.~(\ref{SM_eq:relaxation_Hamiltonian}) in the supplementary material. 
				Since engineered dissipation and the Hamiltonian superoperators are all assumed to perform an autonomous error-corrected quantum computation of order $c$ with respect to natural dissipation, the summations in the expressions of Eqs.~(\ref{eq:L_eff_exp_init}-\ref{eq:L_eff_exp}) initiate only $k=c$ because all lower-order contributions cancelled exactly as a consequence of the applicability of the Lemma.
				
				We notice that $\mathcal{L}_{\textrm{E},H}^{v,w}\mathcal{Q}_{\rm{E}}=\mathcal{Q}_{\rm{E}}\mathcal{L}_{\textrm{E},H}^{v,w}=\mathcal{L}_{\textrm{E},H}^{v,w}$ and $\mathcal{L}_{\textrm{E},H}^{v,w}\mathcal{P}_{\rm{E}}=\mathcal{P}_{\rm{E}}\mathcal{L}_{\textrm{E},H}^{v,w}=0$, thus $\mathcal{L}_{\textrm{E},H}^{v,w}$ is an operator which can be restricted to the relaxation subspace of projected states $\tilde{\rho}=\mathcal{Q}_{\rm{E}}\tilde{\rho}$ of the complementary projector $\mathcal{Q}_{\rm{E}}$. Moreover we can also show that $\mathcal{L}_{\textrm{E},H}^{v,w}$ is also necessarily invertible with its eigenvalues having strictly negative real parts in the  relaxation subspace since  $g$ is chosen so that the bound Eq.~(\ref{SM_eq:relaxation_Hamiltonian}) is satisfied: to prove so we remark that the component  $i (g/R)E_{v,w}\mathcal{Q}_{\rm{E}}$ in $\mathcal{L}_{\textrm{E},H}^{v,w}$ is just an imaginary shift proportional  to the identity in that subspace and thus does not affect the (strictly negative) real parts of the eigenvalues of $\mathcal{L}_{\rm{E}}-ig/R\mathcal{Q}_{\rm{E}}\mathcal{H}\mathcal{Q}_{\rm{E}}=\mathcal{Q}_{\rm{E}}\left(\mathcal{L}_{\rm{E}}-ig/R\mathcal{H}\right)\mathcal{Q}_{\rm{E}}$. As  a consequence, the above defined superopator $\mathcal{F}_{\rm{inv}}^{v,w}$ is indeed well-defined,  and $\mathcal{F}_{\rm{inv}}^{v,w}$  can be seen as the exact inverse of $\mathcal{L}_{\textrm{E},H}^{v,w}$ in the relaxation subspace.  Coming back to the full  Hilbert space, $\mathcal{F}_{\rm{inv}}^{v,w}$ is a pseudo-inverse of $\mathcal{L}_{\textrm{E},H}^{v,w}$, i.e., it satisfies $\mathcal{F}_{\rm{inv}}^{v,w}\mathcal{L}_{\textrm{E},H}^{v,w}\mathcal{F}_{\rm{inv}}^{v,w}=\mathcal{F}_{\rm{inv}}^{v,w}$ and $\mathcal{L}_{\textrm{E},H}^{v,w}\mathcal{F}_{\rm{inv}}^{v,w}\mathcal{L}_{\textrm{E},H}^{v,w}=\mathcal{L}_{\textrm{E},H}^{v,w}$.

				Finally, one remarks that there exists some constants $A,R_0>0$ such that for any $R\geq R_0$ and $|g|\leq \tilde{g}_0 R$ the power series in the expressions of $\mathcal{L}_{\rm{eff}}^{ab}$ in Eq.~(\ref{eq:L_eff_exp}) are all convergent and that the contributions $\mathcal{L}_{\rm{eff}}^{ab}$  ($a,b=0,1$) satisfy $\|\mathcal{L}_{\rm{eff}}^{ab}\|\leq  A/R^c$. Substracting the logical Hamiltonian component,  the  suppression up  to order $c$  of the effective code space Liouvillian $\|\mathcal{L}_{\rm{eff}}-ig\kappa\mathcal{H}_{0}\|<4A/R^c$ confirms the picture of an autonomous error-corrected computation of order $c$.
				
				\subsubsection{Special  case: $H=H_0=0$}
				In absence of an Hamiltonian the expression derived in the previous section are drastically simplified, and one obtains
				
				\begin{eqnarray}
				\mathcal{L}_{\rm{eff}}&=&\kappa\sum_{k=c+1}^{+\infty}\left(\frac{-1}{R}\right)^{k-1}\mathcal{P}_{\rm{E}}\left(\mathcal{L}_{\rm{n}}\mathcal{L}_{\rm{E}}^{\ast}\right)^{k-1}\mathcal{L}_{\rm{n}}\mathcal{P}_{\mathcal{C}}\label{SM_eq:effective-liouvillian_auto_QEC}\\
				\mathcal{T}&=&\sum_{n=0}^{\infty}\frac{(-1)^k}{R^k}\left(\mathcal{L}_{\rm{E}}^{\ast}\mathcal{L}_{\rm{n}}\right)^{k}\mathcal{P}_{\mathcal{C}}\label{SM_eq:T_matrix_auto_QEC},
				\end{eqnarray} 
				
				The quantity $\mathcal{L}_{\rm{E}}^{\ast}=-\int_0^{+\infty}du [e^{\mathcal{L}_{\rm{E}}u}\mathcal{Q}_{\rm{E}}]$ is well-defined (as the projector $Q_{\rm{E}}$ restricts the dynamics to relaxation eigenmodes of $\mathcal{L}_{\rm{E}}$), and is a pseudo inverse of $\mathcal{L}_{\rm{E}}$. As previously, effective dissipation in the logical space is suppressed as $1/R^c$.
				
				We remark in the specific case were we chose the engineered dissipation $\mathcal{L}_{\rm{E}}$ according  to the explicit construction introduced  in Eq.~(\jose{7}) of the main manuscript, the pseudo-inverse  and the various projectors in Eqs.~(\ref{SM_eq:effective-liouvillian_auto_QEC}-\ref{SM_eq:T_matrix_auto_QEC}) have some explicit analytical expressions:
				\begin{eqnarray}
				\mathcal{P}_{\rm{E}}\rho&=&P_{\mathcal{C}}\rho P_{\mathcal{C}}+\sum_{n=1}^{c}\sum_{i_{n}=1}^{p_n}F_{\textrm{E},i_{n}}^{[n]}\rho F_{\textrm{E},i_{n}}^{[n]\dagger}+\sum_{q=1}^{q_{\rm{max}}}F_{\textrm{E},q}^{[\textrm{res}]}\rho F_{\textrm{E},q}^{[\textrm{res}]\dagger}\\
				\mathcal{Q}_{\rm{E}}\rho&=&\rho-\mathcal{P}_{\rm{E}}\rho\\
				\mathcal{L}_{\rm{E}}^{\ast}\rho&=&-2(Q_{\mathcal{C}}\rho P_{\mathcal{C}}+P_{\mathcal{C}}\rho Q_{\mathcal{C}})-Q_{\mathcal{C}}\rho Q_{\mathcal{C}}+\sum_{n=1}^{c}\sum_{i_{n}=1}^{p_n}F_{\textrm{E},i_{n}}^{[n]}\rho F_{\textrm{E},i_{n}}^{[n]\dagger}+\sum_{q=1}^{q_{\rm{max}}}F_{\textrm{E},q}^{[\textrm{res}]}\rho F_{\textrm{E},q}^{[\textrm{res}]\dagger}
				\end{eqnarray}

\end{document}